\documentclass[12pt]{article}
\usepackage{authblk}
\usepackage{hyperref}
\usepackage[top=21truemm,bottom=22truemm,left=24truemm,right=24truemm]{geometry}
\usepackage{setspace}
\normalsize
\usepackage{amsmath,amssymb,mathrsfs,amsbsy,latexsym,amsfonts,amsthm}
\usepackage{physics}

\usepackage[Symbol]{upgreek}

\pdfoutput=1 
\usepackage{graphicx}
\usepackage{color}
\usepackage{comment}
\usepackage{here}
\usepackage[utf8]{inputenc}
\usepackage{tensor}
\usepackage{mathtools}

\newcommand{\nn}{\nonumber\\}
\newcommand{\vx}{\vec{x}}
\newcommand{\vxp}{\vec{x'}}
\newcommand{\vy}{\vec{y}}
\newcommand{\vyp}{\vec{y'}}
\newcommand{\vz}{\vec{z}}
\newcommand{\vzp}{\vec{z'}}

\newcommand{\cA}{\mathcal{A}}
\newcommand{\cO}{\mathcal{O}}
\newcommand{\cH}{\mathcal{H}}

\DeclareMathOperator{\sgn}{sgn}
\DeclareMathOperator{\spn}{Span}
\DeclarePairedDelimiter\floor{\lfloor}{\rfloor}
\usepackage{epsf}

\numberwithin{equation}{section}

\title{\bf Target space entanglement\\in\\quantum mechanics of fermions and matrices}

\author[1, 2]{
Sotaro~Sugishita\thanks{\tt sotaro@eken.phys.nagoya-u.ac.jp}
\vspace{5mm}
}
\affil[1]{\it\normalsize Institute for Advanced Research, 
Nagoya University,
Nagoya, Aichi 464-8601, Japan}
\affil[2]{\it\normalsize Department of Physics, 
Nagoya University,
Nagoya, Aichi 464-8602, Japan}
\date{}

\begin{document}
\maketitle
\begin{abstract}
We consider entanglement of first-quantized identical particles  by adopting an algebraic approach. 
In particular, we investigate fermions whose wave functions are 
given by the Slater determinants, as for singlet sectors of  one-matrix models.
We show that the upper bounds of the general R\'enyi entropies are $N \log 2$ for $N$ particles or an $N\times N$ matrix.
We compute the target space entanglement entropy and the mutual information in a free one-matrix model. 
We confirm the area law: the single-interval entropy for  the ground state scales as $\frac{1}{3}\log  N$ in the large $N$ model.
We obtain an analytical $\mathcal{O}(N^0)$ expression of the mutual information for two intervals in the large $N$ expansion. 
\end{abstract}

\newpage

\setcounter{tocdepth}{2}

\setlength{\abovedisplayskip}{12pt}
\setlength{\belowdisplayskip}{12pt}

\tableofcontents
\newpage
\section{Introduction}

Entanglement entropy in quantum field theories (QFTs) was studied \cite{Bombelli:1986rw, Srednicki:1993im} to understand the black hole entropy known as the Bekenstein-Hawking entropy $S=A/4G_N$.
In the context of the AdS/CFT correspondence, this entropy formula was generalized to the Ryu-Takayanagi formula \cite{Ryu:2006bv, Ryu:2006ef} (see also the covariant generalization \cite{Hubeny:2007xt}): 
Area of an extremal surface in the AdS space is related to the entanglement entropy for a dual CFT on the boundary as  $S^\text{CFT}_{EE}=A/4G_N$ (when the classical Einstein gravity is a good description in the bulk). 
It was proposed in \cite{Faulkner:2013ana} (see also \cite{Engelhardt:2014gca}) that we can generalize this formula to include quantum corrections in the bulk as 
$S^\text{CFT}_{EE}=S^\text{bulk}_{gen}$, where $S^\text{bulk}_{gen}$ is the generalized entropy of a \textit{quantum} extremal surface in the bulk and roughly  given by the area of the quantum surface plus the bulk  entanglement entropy  like $S^\text{bulk}_{gen} \sim \expval{A}/4G_N+S^\text{bulk}_{EE}$.\footnote{Explicit computations of the bulk generalized entropy are done, \textit{e.g.}, in  \cite{Miyagawa:2015sql, Sugishita:2016iel, Belin:2018juv}.}

It was proposed in \cite{Das:2020jhy} that 
the bulk generalized entropy for any region enclosed by codimension-two surfaces, which are not restricted to the extremal surfaces, is given by the \textit{target space} entanglement entropy in the holographically dual description.\footnote{See also \cite{Bianchi:2012ev} where it is also conjectured that the entanglement entropy for general surfaces in quantum gravity is given by the Bekenstein-Hawking formula at the leading order. In the AdS/CFT correspondence, the Bekenstein-Hawking formula for general surfaces are also considered to be dual to  the differential entropy in the boundary theory \cite{Balasubramanian:2013lsa, Myers:2014jia}.} 
Entanglement entropy in QFTs usually means the geometric entropy, \textit{i.e.},  the \textit{base space} entanglement entropy. 
The base space entanglement entropy in a $(1+d)$-dimensional QFT ($d\geq 1)$ is defined by partitioning the $d$-dimensional base space, where quantum fields live, into a subregion and its complement. 
We instead consider the entanglement in the target space. 

We have  an issue when we define the target space entanglement. 
The Hilbert space is generally not tensor-factorized with respect to the target space. 
An alternative method is adopted to resolve this issue \cite{Mazenc:2019ety, Anous:2019rqb}. 
This is based on an algebraic approach \cite{ohya2004quantum} (for reviews see also  \cite{Casini:2013rba, Harlow:2016vwg}).
We usually assume that our total Hilbert space is a  tensor product as $\cH=\cH_A\otimes \cH_{\bar{A}}$ and then consider the entanglement between subsystem  $A$ and $\bar{A}$.
However, total Hilbert spaces sometimes cannot take such simple tensor-factorized forms. 
It means that we cannot define the reduced density matrix by taking a partial trace over a subsystem in the usual manner. 
In the algebraic approach, we instead restrict observables\footnote{See also \cite{Zanardi:2004zz} for a similar idea.} and consider the ``reduced" density matrix associated with the subalgebra generated by the selected observables. 
It allows us to define entanglement in general situations as we will review in section~\ref{sec:review}.
This algebraic approach is used to define entanglement entropies for identical particles \cite{Barnum:2004zz, Balachandran:2013hga, Balachandran:2013cq} and also  in (lattice) gauge theories \cite{Casini:2013rba}.

A typical example is the BFSS model \cite{Banks:1996vh}. 
This is a supersymmetric matrix quantum mechanics describing D0-branes and conjectured to provide a non-perturbative formulation of M-theory.
The notion of the base space is meaningless for the quantum mechanics, that is, $(1+0)$-dimensional QFTs.  
While we cannot divide the zero-dimensional ``base space" into subregions, 
we can the target space. 
The target space entanglement of matrix quantum mechanics is investigated in 
\cite{Das:2020jhy, Das:2020xoa, Hampapura:2020hfg}.


In this paper, we consider a simple matrix quantum mechanics: one-matrix models. 
The dynamical variables are a single $N\times N$ hermitian matrix and the model has $\mathrm{SU}(N)$ symmetry. 
The singlet sector described by the eigenvalues can be mapped to a quantum mechanics of non-relativistic non-interacting fermions \cite{Brezin:1977sv}. 
A two-dimensional string theory can be non-perturbatively formulated by a double scaling large $N$ limit of a  one-matrix quantum mechanics
with a potential containing a quadratic maximum like an inverted harmonic oscillator  \cite{Kazakov:1988ch} (for reviews see, \textit{e.g.},  \cite{Klebanov:1991qa, Polchinski:1994mb}). 
Thus, the one-matrix quantum mechanics is a tractable model of holography. 
The entanglement entropy in the model is computed in \cite{Das:1995vj, Hartnoll:2015fca} based on the second-quantized picture of fermions.
 
The aim of this paper is to develop the target space entanglement of one-matrix quantum mechanics based on the algebraic approach. 
Since the models can be regarded as a system of non-interacting fermions, we investigate the cases where the wave functions are given by the Slater determinants. 
In the algebraic approach, the entanglement entropy consists of the classical part and the quantum one as
$S=S_{cl}+S_q$. 
The classical part $S_{cl}$ represents the classical Shannon entropy for a probability distribution where each probability is assigned to each selection sector.
For the Slater determinant wave functions with $N$ fermions, we show that the maximum value of 
$S_{cl}$ is $\mathcal{O}(\log N)$ at large $N$. 
On the other hand, we show that the max of the total entropy $S_{cl}+S_q$ is linear in $N$.
This linear behavior indicates the volume law of the entropy. 
However, the volume law is not satisfied by specific states such as ground states. 
We indeed confirm that the entanglement entropy for a ground state scales as $\log N$.  
It implies that the area law holds for the ground states of local Hamiltonians as in local QFTs. 

We explicitly confirm that the target space entanglement entropy reduces to the usual entanglement entropy in the second quantized picture. 
The entanglement of non-relativistic free $N$ fermions are already  studied, \textit{e.g.} in \cite{klich2008scaling, Calabrese:2011zzb} (see also \cite{Das:2019qaj}). 
In particular, the large $N$ result of entanglement entropy of the single interval is obtained using a technique developed in \cite{Jin_2004, Calabrese_2010}.
We extend this analysis  to the two-interval case and obtain the leading expression of the general R\'enyi entropy and the mutual information in the large $N$ limit.

This paper is organized as follows: 
In section~\ref{sec:review}, for the sake of completeness, we review the algebraic approach of entanglement and the definition of the (R\'enyi) entropy for first-quantized indistinguishable particles (bosons and fermions). 
In section~\ref{sec:slater}, we especially investigate the case of fermions with the Slater determinant wave functions. 
We find that the target space (R\'enyi) entanglement entropies are given by the independent sum of classical Shannon (R\'enyi) entropies [see eqs.~\eqref{formula_EE}, \eqref{formula_Renyi}].
These formulae are the same as those derived in \cite{klich2008scaling, Calabrese:2011zzb} based on the second quantized picture. 
It means that the target space entanglement agrees with the usual base space entanglement in the second quantized picture. 
We will show this fact explicitly in subsection~\ref{subsec:rederive}. 
We present more details on the second-quantized picture in appendix~\ref{app:2nd}.
In section~\ref{sec:free-1dim}, we compute numerically and analytically the target space entanglement entropy and target space mutual information 
in the case without potentials on a one-dimensional space with the periodic boundary condition. 
To obtain the analytical large $N$ result, we use the Fisher-Hartwig conjecture. The detailed computations are written in appendix~\ref{app:FH}.
We also consider the Dirichlet boundary condition in appendix~\ref{app:diric}.

\section{Review of entanglement entropy for first-quantized indistinguishable particles}
\label{sec:review}

For $(1+d)$-dim QFTs, we consider subregion $A$ and its complement in the base space, and usually assume that the total Hilbert space $\cH$ is a tensor product as $\cH=\cH_A\otimes \cH_{\bar{A}}$. 
We then define the reduced density matrix on $A$ by taking the partial trace over $\cH_{\bar{A}}$. 
This ordinary procedure cannot be applied directly to quantum mechanics, or $(1+0)$-dimensional QFTs. 

The algebraic approach enables us to define the notion of entanglement without such a simple tensor product structure. 
Let us summarize this approach here.
First, we give an algebraic viewpoint of the reduced density matrix for ordinary cases where the total Hilbert space is a tensor product $\cH=\cH_A\otimes \cH_{\bar{A}}$. 
Let $\rho$ be the total density matrix. 
The reduced density matrix $\rho_A :=\tr_{\bar{A}}\rho$ on the subsystem $A$ satisfies $\tr_A (\rho_A \cO_A)=\tr [\rho (\cO_A \otimes 1_{\bar{A}})]$ for any operator $\cO_A \in \mathcal{L}(\cH_{A})$.\footnote{Here $\mathcal{L}(X)$ denotes the set of all linear operators on a vector space $X$.} 
It means that if we introduce the following operator $\rho_\cA \in \mathcal{L}(\cH)$;
\begin{align}
\label{rhocA1}
    \rho_\cA= \rho_A \otimes \frac{ 1_{\bar{A}}}{\mathrm{dim}(\cH_{\bar{A}})},
\end{align}
we have $\tr (\rho_\cA \,\cO)= \tr( \rho \, \cO)$ for any operator $\cO$ in the subalgebra 
$\cA=\mathcal{L}(\cH_{A})\otimes 1_{\bar{A}} \subset \mathcal{L}(\cH)$, which is the set of operators localized on $A$. 
Thus, we can regard $\rho_\cA$ as an effective density matrix for observers who can access only observables in the subalgebra $\cA$.

The above characterization of the reduced density matrix $\rho_\cA$ can be applied to any subalgebra even when the total Hilbert space is not a tensor product.
Let $\cA$ be a subalgebra\footnote{A subalgebra $\cA$ is a subset of $\mathcal{L}(\cH)$ closed under operations such that 
\begin{itemize}
\item $1_\cH\in \cA$,
    \item $\forall x \in \mathbb{C},\,\forall\cO \in \cA \Rightarrow x \cO \in \cA$,
    \item $\forall\cO_1, \cO_2 \in \cA \Rightarrow \cO_1+\cO_2 ,\,\, \cO_1\cO_2 \in \cA$,
    \item $\forall\cO\in \cA \Rightarrow \cO^\dagger\in \cA$,
\end{itemize}
where $1_\cH$ is the identity operator on $\cH$.} in $\mathcal{L}(\cH)$.
We define the reduced density matrix $\rho_\cA$ associated with the subalgebra $\cA$ from the total density matrix $\rho$ as a  positive semi-definite operator in $\cA$ satisfying 
\begin{align}
\label{tr_rhoA}
    \tr (\rho_\cA \,\cO)= \tr( \rho \, \cO)
\end{align}
for any $\cO \in \cA$.
Unlike the above example, a general subalgebra $\cA$ cannot take a tensor-factorized form like  $\mathcal{L}(\cH_{A})\otimes 1_{\bar{A}}$. 
Nevertheless, for any choice of $\cA$, one can show that 
the total Hilbert space can be uniquely decomposed into a direct sum of tensor products
as\footnote{This property can be proved if the Hilbert space  $\cH$ is a finite dimensional space (see, \textit{e.g.}, \cite{Harlow:2016vwg}).
For infinite dimensional Hilbert spaces,  we need a more careful treatment.
Although the Hilbert spaces of particles which we consider this paper are actually infinite dimensional spaces, we will see that the decomposition \eqref{H-dr-sum} is valid in this case. 
In this paper we will not further address the issue of infinite dimensional spaces.} 
\begin{align}
\label{H-dr-sum}
    \cH=\bigoplus_k \cH_{A_k}\otimes \cH_{\bar{A}_k}
\end{align}
such that $\cA$ is tensor-factorized in each sector as
\begin{align}
\label{cA_block}
    \cA= \bigoplus_k  \mathcal{L}(\cH_{A_k})\otimes 1_{\bar{A}_k}.
\end{align}

From this structure we can construct the reduced density matrix $\rho_\cA$ satisfying the property \eqref{tr_rhoA} as follows.
Let $\Pi_k$ be the projection onto the sector $\cH_{A_k}\otimes \cH_{\bar{A}_k}$ in the direct sum \eqref{H-dr-sum}.
We define a real number $p_k$ associated with each sector as
\begin{align}
    p_k:=\tr (\Pi_k \rho \Pi_k).
\end{align}
The set $\{p_k\}$ can be regarded as a probability distribution because we have $0\leq p_k \leq 1$ and $\sum_k p_k=1$. These properties follow from the fact that the total density matrix $\rho$ is positive semi-definite and normalized $\tr \rho=1$.
We also define the restricted density matrix $\rho_k$ on each sector as
\begin{align}
\label{def:rhok}
    \rho_k:=\frac{1}{p_k} \Pi_k \rho \Pi_k
\end{align}
which is normalized as $\tr \rho_k=1$.
$\rho_k$ can be regarded as the density matrix on the sector $\cH_{A_k}\otimes \cH_{\bar{A}_k}$ because it is sandwiched by the projection operators $\Pi_k$.
Since each sector $\cH_{A_k}\otimes \cH_{\bar{A}_k}$ is a tensor product, we can take the partial trace of $\rho_k$ over $\cH_{\bar{A}_k}$ as 
\begin{align}
 \rho_{k,A}:=\tr_{\bar{A}_k}\rho_k.   
\end{align}
Then, the reduced density matrix $\rho_\cA$ associated with $\cA$ is defined as\footnote{This is a formal expression if the dimension of $\cH_{\bar{A}_k}$ is infinite.}
\begin{align}
\label{rhoA_gen}
    \rho_\cA:=\bigoplus_k p_k\, \rho_{k,A} \otimes \frac{ 1_{\bar{A}_k}}{\mathrm{dim}(\cH_{\bar{A}_k})},
\end{align}
which is normalized as $\tr \rho_\cA=1$.
Eq.~\eqref{rhoA_gen} is a generalization of \eqref{rhocA1}. 
The desired property, $\tr (\rho_\cA \,\cO)= \tr( \rho \, \cO)$ for all operators $\cO \in \cA$,  holds because the subalgebra $\cA$ takes the form \eqref{cA_block}. 

We can also define the reduced density matrix $\rho_A$ acting on the space $\cH_A:=\bigoplus_k \cH_{A_k}$ as 
\begin{align}
\label{def:rhoA}
    \rho_A:=\bigoplus_k p_k\, \rho_{k,A}.
\end{align}
The entanglement entropy associated with the subalgebra $\cA$ is defined as the von Neumann entropy of this $\rho_A$;
\begin{align}
\label{def:generalEE}
    S(\rho,\cA)&:=-\tr_{\cH_A} \rho_A\log \rho_A
    \\
    &=-\sum_k p_k \log p_k +\sum_k p_k S_{A_k}(\rho_{k}),
    \label{S_cl_q}
\end{align}
where $S_{A_k}(\rho_{k})$ is the entanglement entropy of subsystem $\cH_{A_k}$ for the $k$-th sector `total' density matrix $\rho_{k}$ on $\cH_{A_k}\otimes \cH_{\bar{A}_k}$,
that is,  $S_{A_k}(\rho_{k})$  is nothing but
the von Neumann entropy of the $k$-th sector reduced density matrix $\rho_{k,A} =\tr_{\bar{A}_k}\rho_k$ as 
\begin{align}
  S_{A_k}(\rho_{k})=S(\rho_{k,A})= - \tr_{A_k} \rho_{k,A} \log \rho_{k,A}.
\end{align}
The entanglement entropy given by \eqref{S_cl_q} consists of two parts. The first term is 
\begin{align}
    S_{cl}(\rho,\cA)=-\sum_k p_k \log p_k
\end{align}
called a \textit{classical} part because it is the Shannon entropy of the classical probability distribution $\{p_k\}$.
The second term is 
\begin{align}
    S_{q}(\rho,\cA)=\sum_k p_k S_{A_k}(\rho_{k})
\end{align}
called  a \textit{quantum} part, which is the expectation value of the entanglement entropies $S_{A_k}(\rho_{k})$ with probabilities $p_k$. This quantum part is also called operationally accessible entanglement entropy \cite{Wiseman_2003, Barghathi:2019oxr}.
Indeed, if the decomposition \eqref{def:rhoA} into each sector is associated with a symmetry such as the projection $\Pi_k$ is that into a specific charged sector of the symmetry, the entropy \eqref{def:generalEE} is nothing but the symmetry resolved entanglement entropy  \cite{Goldstein:2017bua, Bonsignori:2019naz}. 
We will see that this is the case for indistinguishable particles, and the decomposition \eqref{def:rhoA} will be done based on the particle numbers in the subsystem $A$ as done in \cite{lukin2019probing}. 

The R\'enyi entropy is also defined as 
\begin{align}
    S^{(n)}(\rho,\cA)&:=\frac{\log \tr_{\cH_A} \rho_A^n}{1-n}
    =\frac{\log\left(\sum_{k} p_k^n \tr_{A_k} \rho_{k,A}^n\right)}{1-n}.
\end{align}
In the limit $n\to 1$, the entanglement entropy is obtained;
\begin{align}
    \lim_{n\to 1}S^{(n)}(\rho;\cA)=S(\rho; \cA).
\end{align}


We will apply the algebraic definition of entanglement to the quantum mechanics of multi identical particles on a generic manifold $M$.
Although $M$ can be any curved manifold with arbitrary dimension, we will represent it for notational simplicity as one-dimensional flat space, \textit{e.g.}, we will write the integration on a $d$-dimensional $M$ with metric $g$ just as $\int_M dx$ instead of $\int_M d^dx \sqrt{g}$.  
We also use the vector notation $\vx=(x_1,\cdots, x_N)$ to represent the set of coordinates of $N$ particles. 
Each $x_i$ $(i=1,\cdots,N)$ denotes the $d$-dimensional coordinates of a particle on $M$.
The $N \times d$ integrals $\int d^d x_1 \sqrt{g(x_1)}\,\cdots \int d^d x_N \sqrt{g(x_N)} \, \psi(x_1,\cdots,x_N)$ will be denoted by $\int d^N x\, \psi(\vx)$ for brevity.

\subsection{Single-particle system}
\label{subsec:single}
Let us begin with the quantum mechanics of a single particle on a manifold $M$.
Let $\cH^{(1)}$ be the entire Hilbert space. 
For a normalized state $\ket{\psi}\in \cH^{(1)}$, we represent the wave function by $\psi(x)=\braket{x}{ \psi}$, where it is normalized as $\int_M dx |\psi(x)|^2=1$.

We divide $M$ into a subregion $A$ and its complement $\bar{A}$, and consider the entanglement between the two regions $A$ and $\bar{A}$.
The obstacle is that the total Hilbert space $\cH^{(1)}$ does not take a tensor product form with respect to the two regions. 
Thus, we cannot use the ordinary definition of the reduced density matrix, and will take the algebraic approach. 
It is convenient to introduce the projection operators onto the region $A$ and also $\bar{A}$ as 
\begin{align}
\label{def:pia-piba}
    \Pi_A=\int_A dx \ketbra{x}{x}, \quad \Pi_{\bar{A}}=\int_{\bar{A}} dx \ketbra{x}{x}.
\end{align}
The total Hilbert space $\cH^{(1)}$ is decomposed into a direct sum as $\cH^{(1)}=\Pi_A\cH^{(1)}\oplus \Pi_{\bar{A}}\cH^{(1)}$. 
Explicitly, a state $\ket{\psi}$ is decomposed into 
\begin{align}
    \ket{\psi}=\int_A dx\, \psi(x)\ket{x}+\int_{\bar{A}} dx\, \psi(x)\ket{x}.
\end{align}
We write the two projected space as
\begin{align}
    \cH_{1}:=\Pi_A\cH^{(1)}, \quad \cH_{0}:=\Pi_{\bar{A}}\cH^{(1)}.
\end{align}
We thus have 
\begin{align}
    \cH^{(1)}=\bigoplus_{k=0}^1 \cH_{k}.
\end{align}
Note that $\cH_{1}$ is the set of states where the particle number in  region $A$ is one, and $\cH_{0}$ means no particle in  region $A$.
Using the notation in \eqref{H-dr-sum}, we have
\begin{align}
    \cH^{(1)}=\bigoplus_{k=0}^1 
    \cH_{A_k}\otimes \cH_{\bar{A}_k},
\end{align}
where $\cH_{A_1}=\cH_{1},\, \cH_{\bar{A}_0}=\cH_{0}$, 
and $\cH_{A_0}, \cH_{\bar{A}_1}$ are the trivial one-dimensional space $\mathbb{C}$.

A natural subalgebra associated with region $A$ is the set of operators acting on the particle in region $A$. 
We thus define the subalgebra $\cA(A)$ as
\begin{align}
    \cA(A)=\spn\left[ \Bigl\{\ketbra{x}{x'} \Big| x,x'\in A  \Bigr\} \cup \Bigl\{\int_{\bar{A}}dx\ketbra{x}{x}\Bigr\}\right].
    \label{N=1_subalg}
\end{align}
Operators in $\cA(A)$ take the forms
\begin{align}
   \int_A dx dx' O(x,x')\ketbra{x}{x'}+c \int_{\bar{A}}dx\ketbra{x}{x} \qquad (c\in \mathbb{C}).
\end{align}
Note that the subalgebra is tensor-factorized in each sector as
\begin{align}
    \cA(A)=\bigoplus_{k=0}^1  \mathcal{L}(\cH_{A_k})\otimes 1_{\bar{A}_k}.
\end{align}

Since we have specified the subalgebra $\cA(A)$, 
we can define the reduced density matrix $\rho_A$ from a given total density matrix $\rho$ following the definition \eqref{def:rhoA}. 
First, we obtain the probability distribution $\{p_0,p_1\}$ as 
\begin{align}
    p_0=\tr \Pi_{\bar{A}} \rho \Pi_{\bar{A}},
    \quad
    p_1=\tr\Pi_{A} \rho \Pi_{A}.
\end{align}
If $\rho$ is pure as $\rho=\ket{\psi}\bra{\psi}$, we have
\begin{align}
    p_0=\int_{\bar{A}}dx |\psi(x)|^2, \quad p_1=\int_{A}dx |\psi(x)|^2.  
\end{align}
Thus, $p_1$ (or  $p_0=1-p_1$) is the probability that we find the particle in the region $A$ (or $\bar{A}$).
Next, the reduced density matrix in each sector, defined by \eqref{def:rhok}, is given by 
\begin{align}
    \rho_0=\frac{1}{p_0}\Pi_{\bar{A}} \rho \Pi_{\bar{A}},
    \quad
    \rho_1=\frac{1}{p_1}\Pi_{A} \rho \Pi_{A}.
\end{align}
The partial traces over $\cH_{\bar{A}_k}$ are 
\begin{align}
 \rho_{0,A}=\tr_{\bar{A}_0}\rho_0=1, \quad \rho_{1,A}=\tr_{\bar{A}_1}\rho_1=\rho_1.   
\end{align}
We therefore obtain the reduced density matrix $\rho_A$;
\begin{align}
    \rho_A=p_0+p_1\rho_1.
\end{align}

The entanglement entropy of the state $\rho$ associated with the subalgebra $\cA(A)$, which we will represent by $S(\rho; A)$,  is given by
\begin{align}
    S(\rho; A)=-\sum_{i=0}^1 p_i \log p_i -p_1 \tr_{A_1} \rho_1\log \rho_1 .
    \label{eq:SA1}
\end{align}
The first term is the classical part, and the second term $-p_1 \tr_{A_1} \rho_1\log \rho_1$ is the quantum part.
Note that this quantum part vanishes if the original state $\rho$ is pure. 
Indeed, if $\rho$ is pure, $\rho_1$ is also, and then $\tr_{A_1} \rho_1\log \rho_1=0$.
For general mixed $\rho$, the quantum part does not vanish. 

The R\'enyi entropy is
\begin{align}
    S^{(n)}(\rho;A)=\frac{\log\left(
    p_0^n+p_1^n\tr_{A_1} \rho_1^n\right)}{1-n}.
\end{align}
If $\rho$ is pure, the R\'enyi entropy is that of the classical distribution $\{p_0,p_1\}$, that is, 
\begin{align}
     S^{(n)}=\frac{\log\left[
    p_1^n+(1-p_1)^n\right]}{1-n}
\end{align}
where $p_0=1-p_1$.

\subsection{Multi-particle systems of bosons or fermions}
We now consider $N$-particle systems of bosons or fermions. 
Let $\cH^{(N)}$ be the Hilbert space. 
It is obtained by symmetrized or anti-symmetrized of the $N$-fold tensor products of one-particle system $\cH^{(1)}$:
\begin{align}
    \cH^{(N)}=\mathcal{S}_\pm [(\cH^{(1)})^{\otimes N}].
\end{align}
Here $\mathcal{S}_+$ denotes symmetrization for bosons and $\mathcal{S}_-$ does anti-symmetrization for fermions as
\begin{align}
    \mathcal{S}_\pm [\ket{\psi_1}\cdots \ket{\psi_N} ]=\frac{1}{\sqrt{N!}}\sum_{\sigma\in S_N}(\pm)^\sigma 
    \ket{\psi_{\sigma(1)}}\cdots \ket{\psi_{\sigma(N)}} 
\end{align}
where $S_N$ is the symmetric group, and $(\pm)^\sigma$ is a sign function of permutations as  $(+)^\sigma=1$ and $(-)^\sigma=\sgn\sigma$. 
We also introduce the projection operator $P^{\pm}$ as 
\begin{align}
  P^\pm   \ket{\psi_1}\cdots \ket{\psi_N} =\frac{1}{N!}\sum_{\sigma\in S_N}(\pm)^\sigma 
    \ket{\psi_{\sigma(1)}}\cdots \ket{\psi_{\sigma(N)}} .
\end{align}
It is easy to check that $(P^\pm)^2=P^\pm$ and $P^\pm$ is the identity operator on $\cH^{(N)}$.
Arbitrary normalized states $\ket{\psi}$ in $\cH^{(N)}$ can be written as 
\begin{align}
    \ket{\psi}=\int d^Nx \psi(x_1, \cdots, x_N) \ket{x_1, \cdots, x_N} 
\end{align}
with the conditions
\begin{align}
    &\psi(x_{\sigma(1)},\cdots, x_{\sigma(N)})=(\pm)^\sigma \psi(x_1, \cdots, x_N),
    \\
    & \int d^Nx |\psi(x_1, \cdots, x_N)|^2=1.
\end{align}
We will sometimes use the following notation: $\psi(\vx)=\psi(x_1, \cdots, x_N)$, $\psi(\vx_\sigma)=\psi(x_{\sigma(1)},\cdots, x_{\sigma(N)})$, and also $\ket{\vx}=\ket{x_1, \cdots, x_N}$, 
$\ket{\vx_\sigma}=\ket{x_{\sigma(1)},\cdots, x_{\sigma(N)}}$.

We introduce the projection operators onto states where $k$ particles are in the region $A$ and the others in the complement $\bar{A}$ as
\begin{align}
    \Pi_k(A):=\binom{N}{k}P^\pm \left(\Pi_A^{\otimes k}\otimes \Pi_{\bar{A}}^{\otimes (N-k)}\right)P^\pm,
\end{align}
where $\Pi_A, \Pi_{\bar{A}}$ are defined in \eqref{def:pia-piba}.
We have $\Pi_k^2=\Pi_k$ and $\Pi_k \Pi_{k'}=0$ for $k\neq k'$.
We can also show that 
\begin{align}
    \sum_{k=0}^N \Pi_k(A)=P^\pm
\end{align}
by noticing $P^\pm=P^\pm (\Pi_A \oplus \Pi_{\bar{A}})^{\otimes N}P^\pm$ because $(\Pi_A \oplus \Pi_{\bar{A}})^{\otimes N}$ is the identity on $(\cH^{(1)})^{\otimes N}$. 
The projection operators $\{\Pi_0(A), \cdots, \Pi_N(A)\}$ decompose
the total Hilbert space $\mathcal{H}^{(N)}$ into $N+1$ sectors as 
\begin{align}
    \mathcal{H}^{(N)}=\bigoplus_{k=0}^N \mathcal{H}_{k},
\end{align}
where $\mathcal{H}_{k}:=\Pi_k(A) \mathcal{H}^{(N)}$.

We now define the subalgebra $\cA(A)$ associated with the region $A$. 
First, general operators $O$ in $\mathcal{L}(\cH^{(N)})$ can be written as
\begin{align}
    O=\int d^Nx d^N x' O(\vx,\vxp)\ketbra{\vx}{\vxp}
\end{align}
with (anti-)symmetrization $O(\vx_\sigma,\vxp_{\sigma'})=(\pm)^{\sigma\sigma'}O(\vx,\vxp)$.
The projected operators $\Pi_k(A) O \Pi_k(A)$, which can be regarded as operators in $\mathcal{L}(\cH_k)$, are computed as  
\begin{align}
\label{prok_O}
    \Pi_k(A) O \Pi_k(A)= \binom{N}{k}^2 P^\pm \int_A d^k y d^ky' \int_{\bar{A}}d^{N-k}zd^{N-k}z'
    O(\vy,\vz,\vyp,\vzp)\ketbra{\vy,\vz}{\vyp,\vzp}P^\pm,
\end{align}
where $\vy, \vyp$ represent the $k$ components of $\vx, \vxp$ restricted in $A$ as $\vy=(x_1, \cdots, x_k)$, and $\vz, \vzp$ represent the $(N-k)$ components of $\vx, \vxp$ in $\bar{A}$ as $\vz=(x_{k+1}, \cdots, x_N)$.
In general, these operators mix particles in region $A$ and $\bar{A}$.
We then define the subalgebra $\cA_k(A) \subset \mathcal{L}(\cH_k)$ as operators nontrivially acting only on particles in region $A$; 
\begin{align}
    \cA_k(A) :=\spn 
    \Bigl\{ P^\pm \int_{\bar{A}}d^{N-k}z\ketbra{\vy,\vz}{\vyp,\vz} P^\pm\,\Big|\, \vy,\vyp\in A  \Bigr\} .
\end{align}
The subalgebra $\cA(A)\subset \mathcal{L}(\cH^{(N)})$ is defined as the direct sum of $\cA_k(A)$;
\begin{align}
    \cA(A):=\bigoplus_{k=0}^N \cA_k(A).
\end{align}
We can schematically write $\cA(A)$ as the set of operators taking the forms 
\begin{align}
    \cA(A)=\bigcup_{k=0}^N\left\{P^\pm \int_A d^k y d^ky' \int_{\bar{A}}d^{N-k}z\,
    \tilde{O}_{k,A}(\vy,\vyp)\ketbra{\vy,\vz}{\vyp,\vz}P^\pm
    \right\}.
\end{align}
Note that $\Pi_k(A)$ are included in  $\cA(A)$ by taking $O_k(\vy,\vyp)=\binom{N}{k}\delta(\vy-\vyp)$, and thus the identity $P^\pm$ is too.

For the projected operators $O_k\equiv \Pi_k(A) O \Pi_k(A) \in \mathcal{L}(\cH_k)$ taking the form \eqref{prok_O}, we define the partial trace $\tr_{\bar{A}}$ as
\begin{align}
\label{def_trba_k}
    \tr_{\bar{A}}O_k:=\binom{N}{k} \int_A d^k y d^k y'  \int_{\bar{A}}d^{N-k}z \,
   O(\vy,\vz,\vy',\vz) \ketbra{\vy}{\vyp},
\end{align}
while we also define 
\begin{align}
  \tr_A[  \tr_{\bar{A}}O_k]:=\int_A d^k y \mel{\vy}{\tr_{\bar{A}}O_k}{\vy}. 
\end{align}
The binomial factor $\binom{N}{k}$ in \eqref{def_trba_k} represents the ways to choose $(N-k)$ particles  restricted in $\bar{A}$ from $N$ particles and is needed so that the matrix elements are given by
\begin{align}
   \mel{\vy}{\tr_{\bar{A}}O_k}{\vyp} =
   \int_{\bar{A}}d^{N-k}z \, \mel{\vy,\vz}{O_k}{\vyp, \vz} \quad\left(=\binom{N}{k}  \int_{\bar{A}}d^{N-k}z \,
   O(\vy,\vz,\vy',\vz)\right)
\end{align}
for the expression \eqref{prok_O}.
The definition ensures $\tr O_k=\tr_A [\tr_{\bar{A}}O_k]$.

For a given total density matrix $\rho$,  the probability distribution $\{p_k\}$ is obtained
by 
\begin{align}
    p_k=\tr[ \Pi_k(A) \rho \Pi_k (A)].
\end{align}
For pure states, $p_k$ are given by
\begin{align}
\label{p_k:pure}
    p_k=\binom{N}{k}\int_A d^k y \int_{\bar{A}}d^{N-k}z\,|\psi(\vy,\vz)|^2.
\end{align}
$p_k$ is the probability that $k$ particles are in the region $A$ for the wave function $\psi(\vx)$.

The restricted density matrix on each sector $\cH_k$ (see the general definition \eqref{def:rhok}) is
given by 
\begin{align}
    \rho_k=\frac{1}{p_k}\Pi_{k}(A) \rho \Pi_{k}(A).
\end{align}
The reduced density matrix on $A$ is defined by
\begin{align}
    \rho_{k,A}=\tr_{\bar{A}}\rho_k.
\end{align}
More explicitly, for the total density matrix  $\rho$ written in the position basis as
\begin{align}
    \rho=\int d^Nx d^N x' \rho(\vx,\vxp)\ketbra{\vx}{\vxp}
\end{align}
with $\rho(\vx_\sigma,\vxp_{\sigma'})=(\pm)^{\sigma\sigma'}\rho(\vx,\vxp)$, we have
\begin{align}
\label{rhokA:rho}
    \rho_{k,A}=\frac{\binom{N}{k}}{p_k}\int_A d^k y d^k y'
    \int_{\bar{A}}d^{N-k}z \,
   \rho(\vy,\vz,\vy',\vz) \ketbra{\vy}{\vyp}.
\end{align}
The reduced density matrix associated with the subalgebra $\cA(A)$ is given by\footnote{This is a formal expression because $\tr \Pi_k(A)$ is not finite. This is not problematic when we compute the expectation value $\tr (O \rho_{\cA(A)})$ for $O \in \cA(A)$.}
\begin{align}
    \rho_{\cA(A)}:=\bigoplus_{k=0}^N p_k \frac{\binom{N}{k}}{\tr \Pi_k(A)}\Pi_k(A)\left[ \rho_{k,A}\otimes 
   \tr_A \Pi_k (A)
    \right]\Pi_k(A).
\end{align}
One can confirm the desired property $\tr (O \rho)=\tr (O \rho_{\cA(A)})$ for any operator $O \in \cA(A)$.

The reduced density matrix on region $A$ is 
\begin{align}
\rho_A=   \bigoplus_{k=0}^N p_k \rho_{k,A}.
\end{align}
The entanglement entropy is given by 
\begin{align}
\label{cl+q}
    S(\rho; A)= S_{cl}(\rho;A) + S_{q}(\rho;A)
\end{align}
with 
\begin{align}
     S_{cl}(\rho;A)&=-\sum_{k=0}^N p_k \log p_k,
     \label{Scl:N}
     \\
      S_{q}(\rho;A)&= -\sum_{k=0}^N p_k \tr_A \rho_{k,A}\log \rho_{k,A}.
      \label{Sq:N}
\end{align}
Note that $\rho_{0,A}=1$. 
The decomposition \eqref{cl+q} is the same as that done in  \cite{lukin2019probing} based on the particle number conservation. 
The classical part $S_{cl}$ is the entropy for the fluctuation of the particle numbers in the subsystem $A$, and the quantum part $S_q$ is the configurational entanglement entropy which is the weighed sum of the entanglement entropy for each particle-number sector 
\cite{lukin2019probing}. 
\eqref{cl+q} is also regarded as the symmetry resolved entanglement entropy \cite{Goldstein:2017bua, Bonsignori:2019naz} where the conserved charge is now the particle number.

We can also consider the R\'enyi entropy as
\begin{align}
    S^{(n)}(\rho;A)=\frac{\log\left(\sum_{k=0}^N p_k^n \tr_A \rho_{k,A}^n\right)}{1-n}.
\end{align}
For a single-particle system, we have seen that the entanglement entropy and the R\'enyi entropy are just those for the  classical probability distribution $\{p_k\}$ if the state is pure. 
This is not the case for multi-particle systems. 
We generally have quantum contributions even if the state is pure.
The above R\'enyi entropy is also the same as the symmetry resolved  R\'enyi entropy \cite{Goldstein:2017bua, Bonsignori:2019naz} with the conserved charge is the particle number.

\section{Entanglement for the Slater determinants}
\label{sec:slater}
We have presented the general definitions of the reduced density matrix and the (R\'enyi) entanglement entropy for indistinguishable particles in the previous section.
We now explicitly compute them for fermionic particles with the Slater determinant wave functions.
Although the Slater determinant wave functions are typical eigenfunctions for Hamiltonians without multi-body interactions, we here do not assume anything about Hamiltonians and just develop the formula of the (R\'enyi) entanglement entropy for the \textit{given} Slater determinant wave functions.  
It will turn out that the entanglement entropy is given just by a sum of classical Shannon entropy as 
\begin{align}
    S=\sum_{i=1}^N H(\lambda_i).
\end{align}
Here $\lambda_i\, (i=1, \cdots,N)$ are real numbers in the range $0 \leq \lambda_i \leq 1$, which are determined from the given Slater determinant, and 
$H(\lambda)$ is the Shannon entropy of the probability distribution $\{\lambda,1-\lambda\}$ (the Bernoulli distribution),
\textit{i.e.},
\begin{align}
    H(\lambda):=-\lambda \log \lambda -(1-\lambda)\log(1-\lambda).
\end{align}
More generally, we will show that the R\'enyi entropy is given by a sum of that for the $N$ independent Bernoulli distributions as
\begin{align}
    S^{(n)}=\sum_{i=1}^N H^{(n)}(\lambda_i),
\end{align}
where $H^{(n)}(\lambda)$ is the classical R\'enyi entropy of the Bernoulli distribution defined as
\begin{align}
     H^{(n)}(\lambda):=\frac{\log[\lambda^n+(1-\lambda)^n]}{1-n}.
\end{align}
In other words, the (R\'enyi) entanglement entropy is effectively the same as that for $N$ \textit{distinguishable} particles.

In fact, these formulae are already obtained following the usual definition of the \textit{base space} entanglement entropy in the second quantized picture \cite{Klich:2004pb, Calabrese:2011zzb}. It means that the target space entanglement entropy defined in the previous section agrees with the usual base space entanglement entropy in the second quantized picture. We will see this fact explicitly in subsection~\ref{subsec:rederive}.

\subsection{General formulae of entanglement entropy and the R\'enyi entropy}
We consider the following Slater determinant wave function for $N$ fermions:
\begin{align}
\label{Slater}
   \braket{\vx}{\psi} =\psi(\vx)=\frac{1}{\sqrt{N!}}\det(\chi_i(x_j))=\frac{1}{\sqrt{N!}}\sum_{\sigma\in S_N}(-)^\sigma\chi_1(x_{\sigma(1)})\cdots \chi_N(x_{\sigma(N)}),
\end{align}
where $\chi_i(x)$ are the one-body wave functions normalized as
\begin{align}
\label{orthonorm}
    \int_M dx\, \chi_i(x)\chi_j^\ast(x)=\delta_{ij}.
\end{align}

It is convenient to introduce the following $N\times N$ overlap matrices:
\begin{align}
\label{def:X}
    X_{ij}(A)&:=\int_A dx \, \chi_i(x)\chi_j^\ast(x), 
    \\
    \bar{X}_{ij}(A)&:=X_{ij}(\bar{A)}=\int_{\bar{A}} dx \, \chi_i(x)\chi_j^\ast(x)=\delta_{ij}-X_{ij}(A).
\end{align}
Matrix $X(A)$ is generally not diagonal for subregion $A$ because $\chi_i(x)$ are orthonormalized only over the entire region $M$ as \eqref{orthonorm}.
We can diagonalize it by taking an appropriate basis. 
Because $X_{ij}$ is an Hermitian matrix, it is diagonalized by a unitary matrix $U$ as $X_{ij}=(U^{-1}\Lambda U)_{ij}$ where $\Lambda$ is a diagonal matrix whose diagonal components are denoted by $\lambda_i$.
We represent the one-body wave functions in the new basis by $\tilde{\chi}_i(x)$ which are related to the original wave functions as $\tilde{\chi}_i(x)=U_{ij}\chi_j(x)$. 
In terms of the new wave functions, $N$-body wave function $\psi(\vx)$ is given by
\begin{align}
    \psi(\vx)=\frac{1}{\sqrt{N!}\det U}\det(\tilde{\chi}_i(x_j)).
\end{align}
Note that the eigenvalues $\lambda_i$ are  probabilities that we find a particle in region $A$ for the one-body wave functions $\tilde{\chi}_i$; 
\begin{align}
    \lambda_i=\int_A dx \,|\tilde{\chi}_i(x)|^2,
\end{align}
and $\lambda_i$ are in the range $0\leq \lambda_i \leq 1$.

Since the state is specified, we can compute the probability distribution $\{p_k\}$ given by \eqref{p_k:pure} as
\begin{align}
    p_k&=\frac{\binom{N}{k}}{N!}\sum_{\sigma, \sigma'\in S_N}(-)^{\sigma\sigma'}
    \prod_{i=1}^k \left[\int_A d y_i \tilde{\chi}_{\sigma(i)}(y_i)\tilde{\chi}_{\sigma'(i)}^\ast(y_i)\right]
    \prod_{j=1}^{N-k} \left[\int_{\bar{A}}d z_j \tilde{\chi}_{\sigma(j+k)}(z_j)\tilde{\chi}_{\sigma'(j+k)}^\ast(z_j)\right]
    \nn
    &=\sum_{I\in F_k}\prod_{i \in I}\lambda_i \prod_{j \in \bar{I}}(1-\lambda_j)\,, 
    \label{pk:Fk}
\end{align}
where $F_k$ is the set of all subsets of $k$ different integers selected from $\{1,2,\cdots,N\}$.
For example, if $N=3, k=2$, we have $F_2=\Bigl\{\{1,2\}, \{2,3\},\{1,3\}\Bigr\}$. 
For a subset $I \in F_k$, $\bar{I}$ represents the complement: $\bar{I}=\{1,2,\cdots,N\}\setminus I$. 
Let us introduce the following notation
\begin{align}
   \lambda_I:= \prod_{i \in I}\lambda_i\,, 
   \qquad
   \bar{\lambda}_I:=\prod_{j \in \bar{I}}(1-\lambda_j)\,.
\end{align}
For example, if $N=4$ and $I=\{1,2\}$, we have $\lambda_I=\lambda_1\lambda_2$ and $\bar{\lambda}_I=(1-\lambda_3)(1-\lambda_4)$.
In this notation, $p_k$ are simply written as
\begin{align}
    p_k=\sum_{I\in F_k} \lambda_I \bar{\lambda}_I\,. 
\end{align}

The probability distribution $\{p_k\}_{k=0}^N$ is 
the Poisson binomial distribution with  success probabilities $\lambda_1, \cdots, \lambda_N$. 
That is,  $p_k$ is the probability that we find $k$ particles in region $A$ when each particle is found in region $A$ with probability $\lambda_i$.
The classical part of the entanglement entropy \eqref{Scl:N} is the Shannon entropy of this distribution. It is known that,\footnote{The Shannon entropy of the Poisson binomial distribution is a concave function of success probabilities $\lambda_1, \cdots, \lambda_N$ \cite{10.3150/16-BEJ860}.} if we fix $N$ and the mean  success probability $\frac{1}{N}\sum_i \lambda_i\equiv \lambda$, the Shannon entropy is bounded from above \cite{SHEPP1981201, 930936} by that of the binomial distribution $\mathrm{B}(N,\lambda)$, which is the special case of the Poisson binomial distribution with 
$\lambda_1= \cdots= \lambda_N=\lambda$. 
We thus have the upper bound
\begin{align}
    S_{cl}(\lambda_1, \cdots, \lambda_N)\leq S_{cl}[\mathrm{B}(N,\lambda)]. 
\end{align}
The large $N$ behavior of the Shannon entropy of the binomial distribution $\mathrm{B}(N,\lambda)$ is known as
\begin{align}
\label{SBin:largeN}
    S_{cl}[\mathrm{B}(N,\lambda)]=\frac{1}{2}\log[2\pi N\lambda(1-\lambda)]+\frac{1}{2}+\mathcal{O}(1/N).
\end{align}
In fact, by the central limit theorem (or the de Moivre-Laplace theorem), distribution $\mathrm{B}(N,\lambda)$  at large $N$ is approximated well by the normal distribution as
\begin{align}
    p_k \simeq \rho(k):=\frac{1}{\sqrt{2\pi N \lambda(1-\lambda)}}e^{-\frac{(k-N\lambda)^2}{2 n \lambda(1-\lambda)}}.
\end{align}
The Shannon entropy (or the differential entropy\footnote{The differential entropy in information theory is the continuous extension of the Shannon entropy to continuous distributions. Do not confuse it with the differential entropy in holography \cite{Balasubramanian:2013lsa, Myers:2014jia}.})
of this normal distribution reproduces the large $N$ behavior \eqref{SBin:largeN} as
\begin{align}
    -\int^{\infty}_{-\infty}\!\!\! dk\, \rho(k) \log \rho(k)=\frac{1}{2}\log [2\pi N\lambda(1-\lambda)]+\frac{1}{2}.
    \label{cl-bound}
\end{align}
Therefore, at large $N$, the classical part of the entanglement entropy is bounded as\footnote{A similar result is obtained in \cite{klich2008scaling}.
In fact, if we fix the mean and also the variance and suppose that the support of the distribution can be approximated well by the continuous region $(-\infty,\infty)$, the upper bound is given by the entropy of the normal distribution which agrees with the large $N$ behavior.} 
\begin{align}
    S_{cl}(\rho;A)\lesssim \mathcal{O}(\log N).
\end{align}
On the other hand, we will see that the quantum part \eqref{Sq:N} can take a much larger value as $S_q(\rho;A)\lesssim \mathcal{O}(N)$ in general.

In order to obtain the quantum part $S_q(\rho;A)$ given by \eqref{Sq:N}, we compute the reduced density matrices $\rho_{k,A}$ for $k=0,\cdots, N$.
The matrix elements in the position basis are given by
\begin{align}
\label{mel:rhokA}
   \mel{\vy}{\rho_{k,A}}{\vyp}&=\frac{\binom{N}{k}}{p_k} \int_{\bar{A}}d^{N-k}z\,\psi(\vy,\vz)\psi^\ast(\vyp,\vz), \qquad \vy ,\vyp \in A.
\end{align}

Recall that we introduced, around \eqref{pk:Fk}, the set $F_k$, which is the set of all the possible sets of $k$ integers. 
We now introduce $k$-body wave functions associated with the subset $I=\{i_1, \cdots, i_k\}$ in $F_k$ as
\begin{align}
\label{k-wf}
        \psi_I(\vy)=\frac{1}{\sqrt{\lambda_{I}}} \sum_{\sigma\in S_k}\frac{(-1)^\sigma}{\sqrt{k!}} \tilde{\chi}_{i_{\sigma(1)}}(y_1)\cdots\tilde{\chi}_{i_{\sigma(k)}}(y_k).
\end{align}
The wave functions are orthonormalized as 
\begin{align}
    \int_A d^k y \, \psi_I(y) \psi_J^\ast (y)=\delta_{I,J} \qquad \text{for} \quad I, J \in F_k\,,
\end{align}
since $\tilde{\chi}_i$ satisfy $\int_A dy \, \tilde{\chi}_i(y)\tilde{\chi}_j^\ast(y)=\lambda_i \delta_{i,j}$.
Using these $k$-body wave functions, we can find that the matrix elements \eqref{mel:rhokA} are written as
\begin{align}
\label{elem:kbody-A}
    \mel{\vy}{\rho_{k,A}}{\vyp}
     &=\frac{1}{p_k}\sum_{I\in F_k} \lambda_I \bar{\lambda}_I\psi_I(\vy)\psi_I^\ast(\vyp). 
\end{align}
We also have
\begin{align}
    \mel{\vy}{\rho_{k,A}^n}{\vyp}
     &=\frac{1}{p_k^n}\sum_{I\in F_k}( \lambda_I \bar{\lambda}_I)^n\,\psi_I(\vy)\psi_I^\ast(\vyp). 
\end{align}

These expressions mean that the quantum part of the entanglement entropy $S_q(\rho,A)$ is computed as
\begin{align}
S_q(\rho;A)&=    -\sum_{k=0}^N p_k \sum_{I\in F_k} \frac{\lambda_I \bar{\lambda}_I}{p_k} \log \frac{\lambda_I \bar{\lambda}_I}{p_k}
=
\sum_{k=0}^N p_k\log p_k
-\sum_{k=0}^N \sum_{I\in F_k} \lambda_I \bar{\lambda}_I\log (\lambda_I \bar{\lambda}_I).
\end{align}
The first term is the minus of the classical part $S_{cl}=-\sum_{k=0}^N p_k\log p_k$.
The total entanglement entropy is thus given by 
\begin{align}
  S(\rho;A)=S_{cl}+S_q=  -\sum_{k=0}^N \sum_{I\in F_k} \lambda_I \bar{\lambda}_I\log (\lambda_I \bar{\lambda}_I).
\end{align}
Furthermore, we have the identity, which can be easily shown by induction, 
\begin{align}
\label{formula}
      -\sum_{k=0}^N \sum_{I\in F_k} \lambda_I \bar{\lambda}_I\log (\lambda_I \bar{\lambda}_I)= \sum_{i=1}^N H(\lambda_i), \qquad \text{with}\quad  H(\lambda):=-\lambda \log \lambda -(1-\lambda)\log(1-\lambda).
\end{align}
Therefore, we obtain the formula 
\begin{align}
\label{formula_EE}
    S(\rho;A)=\sum_{i=1}^N H(\lambda_i).
\end{align}

The R\'enyi entropy  is also computed as 
\begin{align}
 S^{(n)}(\rho;A)  =\frac{\log\left(\sum_{k=0}^N p_k^n \tr_A \rho_{k,A}^n\right)}{1-n} =\frac{\log\left(\sum_{k=0}^N \sum_{I\in F_k} (\lambda_I\bar{\lambda}_I)^n \right)}{1-n},
\end{align}
because we have 
\begin{align}
    \tr_A \rho_{k,A}^n=\sum_{I\in F_k}\left(\frac{\lambda_I \bar{\lambda}_I}{p_k}\right)^n.
\end{align}
Similarly to \eqref{formula}, the following identity holds
\begin{align}
    \sum_{k=0}^N \sum_{I\in F_k} (\lambda_I \bar{\lambda}_I)^n=\prod_{i=1}^N [\lambda_i^n+(1-\lambda_i)^n].
\end{align}
Therefore, we obtain the formula for the R\'enyi entropy,
\begin{align}
\label{formula_Renyi}
    S^{(n)}(\rho;A)=\sum_{i=1}^N H^{(n)}(\lambda_i),
\end{align}
where 
\begin{align}
     H^{(n)}(\lambda):=\frac{\log[\lambda^n+(1-\lambda)^n]}{1-n}.
\end{align}
It is clear that the entanglement entropy \eqref{formula_EE} is obtained in the limit $n\to 1$ in \eqref{formula_Renyi}.

Recall that $\lambda_i$ is the probability that we find a particle in region $A$ for the one-body wave function $\tilde{\chi}_i(x)$, and then the (R\'enyi) entanglement entropy for this single-particle system is given by the Shannon entropy $H^{(n)}(\lambda_i)$ as we have seen in section~\ref{subsec:single}.
Thus, the formulae \eqref{formula_EE} and \eqref{formula_Renyi} indicate that the (R\'enyi) entropy is effectively the same as the sum of the entropies of $N$ distinguishable particles.
More explicitly, the (R\'enyi) entanglement entropy is the same as that of the following density matrix:
\begin{align}
    \rho_\text{eff}=\bigotimes_{i=1}^N \rho^{(i)}, \qquad 
     \rho^{(i)}:=
     \begin{pmatrix}
     \lambda_i &0\\
     0&1-\lambda_i
     \end{pmatrix}.
     \label{dens:indep}
\end{align}
This structure is explicit in the second-quantized picture (see appendix~\ref{app:2nd}).

We can obtain the upper bound of the (R\'enyi) entropy from the formulae \eqref{formula_EE} and \eqref{formula_Renyi}.
Since the R\'enyi entropy function $ H^{(n)}(\lambda)$ takes the maximum value $\log 2$ at $\lambda=1/2$, 
the R\'enyi entropy is bounded independently of $n$ as
\begin{align}
\label{Renyi:bound}
    S^{(n)}(\rho;A)\leq N\log 2.
\end{align}
This means that the quantum part of the entanglement entropy can be $\mathcal{O}(N)$ unlike the classical part which is bounded as $  S_{cl}\lesssim \mathcal{O}(\log N)$ as shown in \eqref{cl-bound}.

The bound \eqref{Renyi:bound} means that the (R\'enyi) entropy is always finite  if $N$ is finite unlike QFTs where the entanglement entropy is generally UV divergent. 
The bound is also independent of the dimension of the  target space $M$ where particles live. 
This finiteness  is similar to a $N$ qubit system. 
However, it should be remarked that the dimension of the Hilbert space of the $N$ particle system is infinite because particles live in a continuum space, while the dimension of the $N$ qubit system is a finite value $2^N$.

The maximum value $N\log 2$ in the bound  \eqref{Renyi:bound} can be interpreted as follows; 
This maximum value is realized when all $\lambda_i$ are equal to $1/2$. 
It means that all the possible $2^N$ configurations where $N$ particles are assigned to either $A$ or $\bar{A}$ have the equal probability $1/2^N$. The entropy is $S=N\log 2$. 

We should also comment that the upper bound is too generic like the volume law of entropy in QFTs. 
It is known for QFTs that 
the volume law of entropy is satisfied by generic states but not by 
physically interesting states like the ground states of local Hamiltonians (or spin systems with local interactions).
They follow the area law.  
This is also the case for quantum mechanics.
We will see that the entanglement entropy for the ground state of free fermions in a finite one-dimensional region behaves as $S\sim \mathcal{O}(\log N)$, not $\mathcal{O}(N)$. 
This is a counterpart of the area law in quantum mechanics of particles.

\subsection{Consistency with  the second-quantized picture}
\label{subsec:rederive}
The formula of entanglement entropy \eqref{formula_EE}  can be written in terms of the $N\times N$ overlap matrix $X$,  defined in \eqref{def:X}, as 
\begin{align}
     S(\rho;A)=-\tr [X\log X+(1_N-X)\log(1_N-X)].
     \label{SformX}
\end{align}
It resembles the formula (see \cite{peschel2003calculation, Casini:2009sr, Song:2011gv}) of free fermions on lattice (or for continuum field theories) given by
\begin{align}
\label{fomla:2nd}
    S(\rho;A)=-\tr_A [G\log G+(1_L-G)\log(1_L-G)].
\end{align}
Here we consider the entanglement entropy for the subsystem $A$ containing $L$ lattice sites, and $G$ is the correlation matrix restricted on $A$ as 
\begin{align}
\label{corr-lat}
G_{ab}:=\tr (c^\dagger_a c_b \, \rho)=\tr_A (c^\dagger_a c_b\, \rho_A), \qquad a,b \in A,
\end{align}
where $c^\dagger_a, c_a$ are the creation and annihilation operators of fermions at site $a$. 
$G$ and $1_L$ are $L\times L$ matrices. 
Their size is infinite if we consider the continuum space $L \to \infty$ as in the present paper, while the size of the overlap matrix $X$ is finite.
Thus, for general cases, the formulae \eqref{SformX}  and \eqref{fomla:2nd} are different. 
Nevertheless, they give the same result, if the number of particles is fixed to $N$ and the state is given by the Slater determinant.
We will show that \eqref{fomla:2nd} indeed reduces to \eqref{SformX}.
Indeed, the formula \eqref{SformX} is already obtained in this way in \cite{Klich:2004pb, Calabrese:2011zzb} although in these references  the  algebraic approach and the target space entanglement are not mentioned.\footnote{The equivalence of the first and second quantization is also shown in 
\cite{Mazenc:2019ety, Das:2020jhy}.
} 
Other detailed computations on the second quantization are presented in appendix~\ref{app:2nd}.

In the second-quantized picture, $N$-body pure states can generally be written as 
\begin{align}
    \ket{\psi}=\frac{1}{\sqrt{N!}}\int d^N x\, \psi(x_1,\cdots, x_N)c^\dagger(x_1)\cdots c^\dagger(x_N)\ket{0}
\end{align}
by acting fermionic ladder operators satisfying 
\begin{align}
\label{ccd:cnt}
    \{c(x),c^\dagger(y)\}=\delta(x-y)
\end{align}
on the Fock vacuum $\ket{0}$.
We introduce the two-point correlation function
\begin{align}
    G(x;y):=\bra{\psi} c^\dagger (x) c(y)\ket{\psi},
\end{align}
which is computed as
\begin{align}
   G(x;y)=N\int d^{N-1} w\, \psi^\ast (x,w_1,\cdots, w_{N-1}) \psi(y,w_1,\cdots, w_{N-1}).
\end{align}

If the wave function $\psi$ is given by the Slater determinant as \eqref{Slater}, $G$ is further simplified as
\begin{align}
   G(x;y)
=\sum_{i=1}^N\chi_{i}^\ast(x)\chi_{i}(y).
\end{align}
We are also able to compute 
multi-point functions using  Wick's theorem. The proof of Wick's theorem is given in appendix~\ref{app:2nd}.
This Wick's theorem ensures the formula \eqref{fomla:2nd}.

Note that, in our continuum space case, the label of sites  $a,b$  in \eqref{corr-lat} correspond to coordinates $y_1, y_2 \in A$, and the identity matrix $1_L$ is now the delta function $\delta(y_1,y_2)$.
In this matrix notation, the matrix product of $G$ restricted on $A$ is given by
\begin{align}
\label{G:product}
    G^2(y_1;y_2):=\int_A dy\, G(y_1;y) G(y;y_2)=\sum_{i,j=1}^N 
    \chi_{i}^\ast(y_1) X_{ij}\chi_{j}(y_2),
\end{align}
which means $\tr_A G^k=\tr X^k$. 
Using this fact, we can show that the RHS of \eqref{fomla:2nd} becomes
\begin{align}
     S(\rho;A)=-\tr [X\log X+(1_N-X)\log(1_N-X)].
\end{align}

It is hard to numerically compute the RHS of \eqref{fomla:2nd} when the number of lattice sites $L$ is large because we have to treat $L \times L$ correlation matrix $G$. 
The equivalence of \eqref{SformX}  and \eqref{fomla:2nd} means that we do not have to compute the correlation matrix if we know the form of the Slater determinant. 
The direct use of the formula \eqref{SformX} is more efficient for $N \ll L$, 
in particular, when we consider finite particles on a continuum space like one-matrix quantum mechanics with a finite rank.

\section{Entanglement for the ground state of \texorpdfstring{$N$}{N} free fermions in one-dimensional space}\label{sec:free-1dim}
We will explicitly compute the (R\'enyi) entanglement entropy for the ground state of $N$ non-interacting fermions in one-dimensional space without a potential.
We consider a finite region to make the spectrum discrete. 
In this section, we take a circle $(-L/2 \leq x \leq L/2)$  with length $L$ as the entire space. 
In appendix~\ref{app:diric},
we also consider a finite interval with the Dirichlet boundary condition.
The result for large $N$ is similar to the periodic case if the subregion does not touch the boundary of the interval. 
We can think that the result obtained here is independent of the boundary condition. 
Thus, we can regard the result as that for the matrix quantum mechanics. 

Energy eigenfunctions of a single particle on the circle are given by
\begin{align}
\label{free_chi}
    \chi_i(x)=\frac{1}{\sqrt{L}}e^{\frac{2\pi i}{L} n_i x},
\end{align}
where $n_i$ are integers as 
\begin{align}
    n_1=0,\, n_2=-1,\, n_3=1,\, n_4=-2,\, n_5=2,\, \cdots,
\end{align}
which can be written using the floor function as $ n_i=(-1)^{i+1}\floor*{\frac{i}{2}}$.

To avoid the degeneracy, we suppose that the total particle number $N$ is odd. Then, the ground state of the $N$-body system is unique. 
The wave function is given by the Slater determinant
\begin{align}
    \psi(\vx)=\frac{1}{\sqrt{N!}}\sum_{\sigma\in S_N}(-)^\sigma\chi_1(x_{\sigma(1)})\cdots \chi_N(x_{\sigma(N)}).
\end{align}

\subsection{Single interval}
We first consider the case where subregion $A$ is an interval; $A=(a L,b L)$ $(-1/2 \leq a<b\leq 1/2)$.
This is considered in \cite{Calabrese:2011zzb} (based on the second-quantized picture), and the large $N$ result is  analytically obtained. 
In fact, the problem to find the eigenvalues of the overlap matrix $X(A)$ defined in \eqref{def:X} for a single interval is exactly the same as finding the eigenvalues of the correlation matrix for an $N$-spin subsystem of the XX spin chain model. The single-interval entanglement entropy of the spin chain is analytically computed for large $N$ in \cite{Jin_2004, Calabrese_2010}.

The overlap matrix $X(A)$ is a function of $a,b$: 
\begin{align}
\label{eq:Xab}
    X_{ij}(a,b)&=\int^{b L}_{aL}\!\!dx\,\chi_i(x) \chi_j^\ast(x)
    =\frac{e^{2\pi i (n_i-n_j)b}-e^{2\pi i (n_i-n_j)a}}{2\pi i (n_i-n_j) }.
\end{align}
Note that $X(a,b)$ does not depend on the length of the circle $L$. In addition, 
the eigenvalues of $X(a,b)$ are invariant under translation $a\to a+c, b\to b+c$, because we have 
\begin{align}
    X(a+c,b+c)=U(c)X(a,b)U^\dagger(c),
\end{align}
where $U(c)$ is a diagonal unitary matrix with the components $U_{ij}(c)=e^{2\pi i n_ic}\delta_{ij}$.
Thus, the (R\'enyi) entanglement entropy for a single interval is translation invariant and depends only on $(b-a)$.

\paragraph{Illustration}
For an illustration, let us begin with $N=3$ and consider the case where the subregion $A$ is a half region of the circle, $A=(- L/2, 0)$.

The wave functions $\tilde{\chi}_i(x)$ diagonalizing the overlap matrix $X$ are explicitly computed as
\begin{align}
    \tilde{\chi}_1(x)&=\frac{1}{\sqrt{2}}[\chi_2(x)+\chi_3(x)]=\frac{\sqrt{2}}{\sqrt{L}}\cos\left(\frac{2\pi x}{L}\right),
    \\
    \tilde{\chi}_2(x)&=\frac{1}{\sqrt{2}}\chi_1(x)+\frac{i}{2}[-\chi_2(x)+\chi_3(x)]=\frac{1}{\sqrt{2L}}-\frac{1}{\sqrt{L}}\sin\left(\frac{2\pi x}{L}\right),
    \\
     \tilde{\chi}_3(x)&=\frac{1}{\sqrt{2}}\chi_1(x)-\frac{i}{2}[-\chi_2(x)+\chi_3(x)]=\frac{1}{\sqrt{2L}}+\frac{1}{\sqrt{L}}\sin\left(\frac{2\pi x}{L}\right).
\end{align}
They are orthogonal to each other on the half space as
\begin{align}
    \int^{0}_{-\frac{L}{2}}\!\!\!dx\, \tilde{\chi}_i(x)\tilde{\chi}^\ast_j(x)=\lambda_i \delta_{ij}
\end{align}
with $\lambda_1=\frac{1}{2}$, $\lambda_2=\frac{1}{2}+\frac{\sqrt{2}}{\pi}\sim 0.95$, $\lambda_3=\frac{1}{2}-\frac{\sqrt{2}}{\pi}\sim 0.05$.
$\tilde{\chi}_1(x)$ is an even function and thus the probability that we find a particle in region $A$ is $\lambda_1=\frac{1}{2}$.
$\tilde{\chi}_2(x)$ and $\tilde{\chi}_3(x)$ are almost localized in region $A$ and $\bar{A}$, respectively.

Since we have obtained the probabilities $\lambda_i$, the R\'enyi entropy is computed as
\begin{align}
    S^{(n)}(N=3)=\sum_{i=1}^3 H^{(n)}(\lambda_i)=\log 2+\frac{2}{1-n}\log\left[\left(\frac{1}{2}+\frac{\sqrt{2}}{\pi}\right)^n+\left(\frac{1}{2}-\frac{\sqrt{2}}{\pi}\right)^n\right].
\end{align}
In particular, the entanglement entropy is $S(N=3)\sim 1.09$.

\begin{figure}[H]
\centering
\includegraphics[width=8cm]{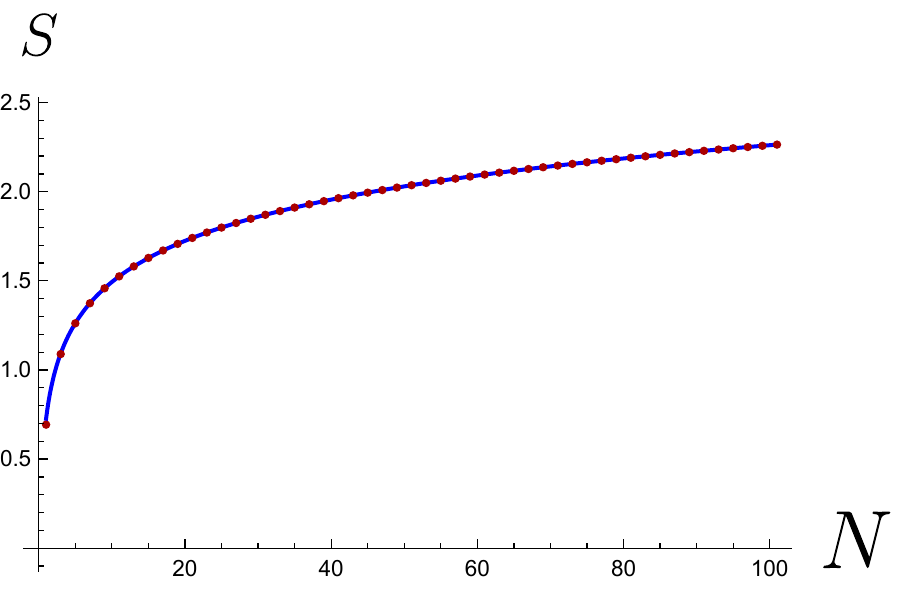}
\caption{Entanglement entropy for the half region. The red dots are the entanglement entropy $S(N)$ for $N=1,3, \cdots, 101$. The blue curve represents the large $N$ result \eqref{EE:sing-large} with $r=1/2$.}
\label{fig:half}
\end{figure}

\paragraph{Large $N$ results}
We can numerically compute the entanglement entropy for arbitrary $N$ by finding the  eigenvalues $\lambda_i$ of $N\times N$ overlap matrix $X$.
The results for the half region with $N=1, 3, \cdots,101$ are shown in
Fig.~\ref{fig:half}.

We can also analytically obtain the asymptotic large $N$ results for any single interval \cite{Jin_2004, Calabrese_2010}. 
Because of  the translational invariance, we can move the center of the interval to the origin. 
Hence we consider the interval $I_1=(-r L/2, r L/2)$ with arbitrary $r$. Then,  by permuting the indices appropriately (see \eqref{appc:Xsing}), we can find that the matrix  \eqref{eq:Xab} can be written as 
\begin{align}
\label{single-Tmat}
    X_{jk}(I_1)=\frac{\sin [\pi(j-k)r]}{\pi(j-k)}.
\end{align}
This is a Toeplitz matrix. 
This Toeplitz matrix also appears in the correlation matrix \eqref{corr-lat} for the tight binding model (which is equivalent to the XX spin chain model\footnote{The parameter $r$ in \eqref{single-Tmat} is related to the Fermi energy in the tight binding model and to the magnetic field in the XX model.}) \cite{Latorre:2003kg, Jin_2004}, and 
the asymptotic large $N$ behavior of the determinant of the Toeplitz matrix can be obtained \cite{Jin_2004, Calabrese_2010} if we use the Fisher-Hartwig conjecture \cite{wu1966theory, fisher1969toeplitz}. The detailed computations are given in \cite{Jin_2004, Calabrese_2010}. We also give computations for the case of two intervals using the Fisher-Hartwig conjecture in appendix \ref{app:FH}. 
Here we just write the result of the (R\'enyi) entanglement entropy for the interval $I_1=(-r L/2, r L/2)$;
\begin{align}
S^{(n)}(I_1) \sim \frac{1}{6}\left(1+\frac{1}{n}\right) \log  [2N \sin(\pi r)]
+\Upsilon_n ,
\end{align}
where we only kept the leading part $\mathcal{O}(\log N)$ and the  subleading part  $\mathcal{O}(N^0)$. The last term $\Upsilon_n$ denotes a real number defined by the following integral: 
\begin{align}
\label{def:Upn}
  \Upsilon_n:=\frac{n
   }{i(1-n)}\int^{\infty}_{-\infty}dw [\tanh(\pi n w)-\tanh(\pi w)]\log \frac{\Gamma\left(\frac{1}{2}+i w\right)}{\Gamma\left(\frac{1}{2}-i w\right)}.
\end{align}
In particular, the entanglement entropy is
\begin{align}
\label{EE:sing-large}
S(I_1)\sim \frac{1}{3} \log  [2N \sin(\pi 
r)]
+\Upsilon_1 
\end{align}
with 
\begin{align}
    \Upsilon_1=i\int^{\infty}_{-\infty}dw \frac{\pi w}{\cosh^2(\pi w)}\log \frac{\Gamma\left(\frac{1}{2}+i w\right)}{\Gamma\left(\frac{1}{2}-i w\right)}\sim 0.495018.
\end{align}

This asymptotic result fits well with the numerical results as shown in Fig.~\ref{fig:half}.
For example, if  the subspace is  a half region, a function of the form $S=\frac{c}{3} \log (2 N)+d$ gives us a good fit\footnote{The fit using the data for $N=1, \cdots, 101$ indicates $c=1.0082$ and $d=0.4824$. 
The fit using a large $N$ region, $N=61, \cdots, 101$, indicates $c=1.0000$ and $d=0.4950$.} even for small $N$. 

The logarithmic behavior in $N$ shows that the entanglement entropy for the ground state is very small 
compared to the maximum value $N \log 2$.
This is similar to the area law of entanglement entropy in local QFTs. 
Indeed, as we mentioned above, the result \eqref{EE:sing-large} is the same as that in the XX model which is described in the continuum limit by  a conformal field theory ($c=1$ free fermions).
The CFT predicts that the entanglement entropy  for an interval with length $r L$  in the circle with length $L$ is given by (see, \textit{e.g.}, \cite{Calabrese:2004eu})
\begin{align}
    S(I_1)=\frac{1}{3}\log\left[\frac{L}{\delta}\sin(\pi r)\right]+c'
\end{align}
where $\delta$ is a UV regulator and $c'$ is a non-universal constant depending on the UV regularization. 
Hence, we can interpret $N$ as a UV cutoff like $N\propto \delta^{-1}$. 

\paragraph{Bosons} Let us also comment on the result for bosons. 
The ground state wave function of $N$ bosons is given by a constant function,
\begin{align}
    \psi_{0}^\text{boson}(\vx)=\frac{1}{L^\frac{N}{2}}.
\end{align}
The probabilities $p_k$ for subregion $A$ with the total length $r L$ are\footnote{Here $A$ is not restricted to a single interval. It can be any union of multiple intervals.}
\begin{align}
    p_k=\binom{N}{k} r^k
\left(1-r\right)^{N-k}.
\end{align}
Thus, the classical part of the entanglement entropy is given by the Shannon entropy for the binomial distribution $\mathrm{B}(N,\ell/L)$.
The reduced density matrix $\rho_{k,A}$ for each sector is pure, and then the quantum part of the entanglement entropy vanishes $S_q=0$.
The total entanglement entropy is
\begin{align}
    S(\rho;A)=S_{cl}[\mathrm{B}(N,r)].
\end{align}
The large $N$ behavior is 
\begin{align}
     S(\rho;A)=\frac{1}{2}\log \left[2\pi N r
\left(1-r\right)\right]+\frac{1}{2}+\mathcal{O}(1/N).
\end{align}
The boson system also has a logarithmic behavior of $N$ but the coefficient is different from that of fermions.

\subsection{Two intervals and mutual information}

We next consider two intervals $I_1=(a_1 L,b_1 L)$, $I_2=(a_2 L,b_2 L)$ which are not overlapped.
The matrix $X$ in \eqref{def:X} satisfies 
\begin{align}
    X_{ij}(I_1\cup I_2)=X_{ij}(I_1)+X_{ij}(I_2).
\end{align}
In general, two matrices $X(I_1)$ and $X(I_2)$ cannot be diagonalized simultaneously.
Hence the mutual information
\begin{align}
  I(I_1;I_2):=  S(I_1)+S(I_2)-S(I_1\cup I_2)
\end{align}
and more generally the R\'enyi mutual information 
\begin{align}
  I^{(n)}(I_1;I_2):=  S^{(n)}(I_1)+S^{(n)}(I_2)-S^{(n)}(I_1\cup I_2)
\end{align}
can be nonzero.
We consider the case that  the two intervals have the same length $r L$, and take the parameters as $(a_1,b_1)=(-d/2-r/2,-d/2+r/2)$, $(a_2,b_2)=(d/2-r/2,d/2+r/2)$.
Note that the distance between the centers of the intervals is $\mathrm{min}(d L, (1-d)L)$.
The condition that $I_1$ and $I_2$ do not  share a region is $r\leq d \leq 1-r$.

The R\'enyi entropy for the two intervals, $S^{(n)}(I_1\cup I_2)$, can be computed in the large $N$ limit using the Fisher-Hartwig conjecture like the single interval case. 
The computations are given in the appendix \ref{app:FH}. 
The result up to the subleading order $\mathcal{O}(N^0)$ is as follows: 
\begin{align}
\label{twointEE}
        S^{(n)}(I_1\cup I_2)\sim  \frac{1}{6}\left(1+\frac{1}{n}\right)\left[2\log[2N\sin (\pi r)]+\log\frac{ \sin[\pi (d+r)] \sin[\pi (d-r)]}{\sin^2(\pi d)}\right] +2\Upsilon_n,
\end{align}
where $\Upsilon_n$ is a constant defined by \eqref{def:Upn}.

Thus, the (R\'enyi) mutual information in the large $N$ limit is 
\begin{align}
\label{min_largeN}
    I^{(n)}(I_1;I_2)\sim \frac{1}{6}\left(1+\frac{1}{n}\right)\log\frac{\sin^2(\pi d)}
    { \sin[\pi (d+r)] \sin[\pi (d-r)]}.
\end{align}
In particular, the mutual information $(n=1)$ is given by
\begin{align}
\label{mi_largeN}
    I(I_1;I_2)\sim \frac{1}{3}\log\frac{\sin^2(\pi d)}
    { \sin[\pi (d+r)] \sin[\pi (d-r)]}.
\end{align}
The (R\'enyi) mutual information is finite even in the large $N$ limit, while 
the (R\'enyi) entanglement entropy diverges as $\mathcal{O}(\log N)$. This reflects the UV finiteness of the mutual information in QFTs.
Eq.~\eqref{min_largeN} agrees with the result in \cite{Calabrese:2004eu} (see also comments on the Calabrese-Cardy result \cite{Calabrese:2004eu} in \cite{Furukawa:2008uk, Calabrese:2009ez}).

\begin{figure}[H]
\centering
\includegraphics[width=8cm]{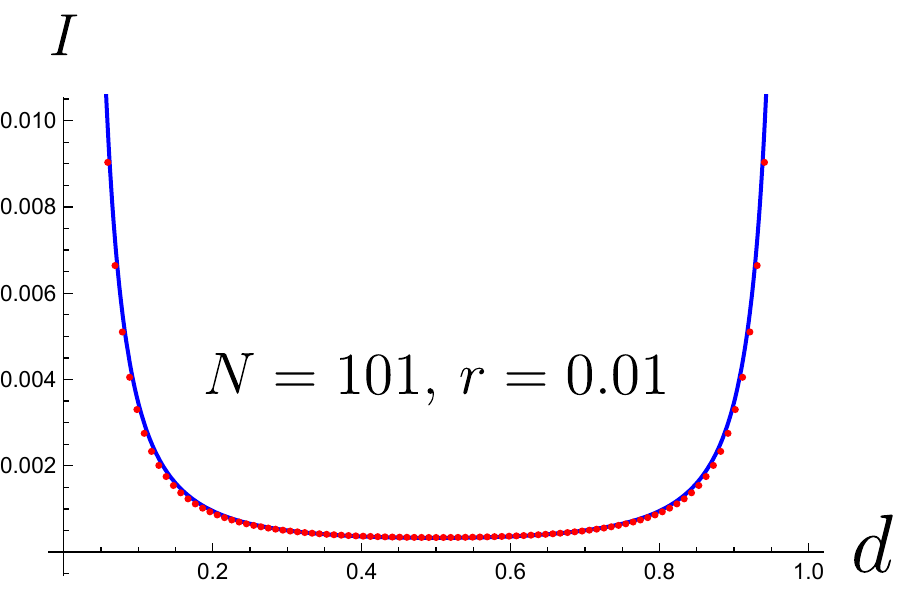}
\caption{Mutual information for two intervals. We take $N=101$ and set the parameter $r$ as $r=0.01$ (length of the intervals is $rL$). The red dots represent  the mutual information for some values of $d$. The blue curve represents the large $N$ result \eqref{mi_largeN}.}
\label{fig:mi}
\end{figure}

In Fig.~\ref{fig:mi}, 
we show the numerical results of the mutual information obtained by the direct diagonalization of  $X_{ij}(I_1\cup I_2)$, and  compare them with the large $N$ result \eqref{mi_largeN}.


\section{Concluding remarks}
We have investigated the target space entanglement, by the algebraic approach, for fermions with the Slater determinant wave functions, which is the same as those of singlet sectors in one-matrix quantum mechanics. 
The entanglement entropy is given by eq.~\eqref{formula_EE}, 
and the R\'enyi entropy is \eqref{formula_Renyi}.
We have shown that the classical part of the entanglement entropy scales as $\mathcal{O}(\log N)$ but the quantum part is generally linear in $N$. 
This is the volume law of entropy for general states. 
However, the entropy for the ground state follows the area law such that it behaves as $\mathcal{O}(\log N)$. 
We have confirmed the area law for free fermions without potentials on a periodic circle. 
The computation of the entanglement entropy for the single interval is exactly the same as that in the XX spin chain. 
The leading term of the entropy is given by 
$S\sim \frac{1}{3}\log N$ at large $N$. 
Unlike QFTs, the entropy is finite (if $N$ is finite), and $N$ plays a role of UV cutoff.

We have also considered the target space R\'enyi entropy and the mutual information for two intervals. 
The large $N$ expression is obtained as \eqref{twointEE} and \eqref{min_largeN}. 
It will be interesting to consider the subleading corrections in the large $N$ expansion (the corrections for the single interval case is considered in \cite{Calabrese_2010}).

We have not considered the one-matrix model dual to a two-dimensional string theory although the formulae 
\eqref{formula_EE}, \eqref{formula_Renyi} can be applied in this case.
In order to consider this model, we have to add a potential and find the wave functions. 
This procedure is done in \cite{Das:1995vj, Hartnoll:2015fca} at the leading order in the large $N$ limit based on the second-quantized picture. 
Since the formula \eqref{formula_EE} is suited for finite $N$, 
it may be interesting to numerically compute the entanglement entropy for finite $N$ and compare the results with the large $N$ ones. 

Extensions to multi-matrix models are important future directions. 
The algebraic approach could be more useful in the directions. However, since the multi matrices in general cannot be diagonalized simultaneously, there are difficulties to define the subalgebra corresponding to a subregion in the target space. 
Two gauge invariant definitions of the subalgebra are proposed in \cite{Das:2020xoa} with the discussions on the ambiguities. 
One of them is argued to be preferable in \cite{Hampapura:2020hfg}.
Anyway, we have not understood well the target space entanglement of the BFSS model and other multi-matrix models, e.g., the BMN model \cite{Berenstein:2002jq}, so far. 
Many things remain to be explored.

\section*{Acknowledgement}
The author is grateful to Sumit Das for reading an early draft of this paper and informing him of the important references.
The author acknowledges  support from JSPS KAKENHI Grant Number JP 21K13927.
\appendix
\section{Second quantization, Wick's theorem and modular Hamiltonian}\label{app:2nd}
In this appendix, we present more detailed computations in the second-quantized picture. 
We have introduced fermionic ladder operators $c(x), c^\dagger(x)$ in \eqref{ccd:cnt}. They can be expanded as 
\begin{align}
\label{mode:c}
    c(x)=\sum_{i\,:\,\text{all}}\chi_i(x) c_i.
\end{align}
Here, the label $i$ in the sum runs over all modes, not restricted to $i=1, \cdots, N$, such that $\chi_i(x)$ are complete orthogonal modes on the entire space $M$ as 
\begin{align}
    \int_M dx\, \chi_i(x) \chi^\ast_j(x)=\delta_{ij}, \qquad 
    \sum_{i\,:\,\text{all}}\chi_i(x) \chi^\ast_i(y)=\delta(x,y).
\end{align}

Then, the second-quantized state 
\begin{align}
    \ket{\psi}=\frac{1}{\sqrt{N!}}\int d^N x\, \psi(x_1,\cdots, x_N)c^\dagger(x_1)\cdots c^\dagger(x_N)\ket{0}
\end{align}
with the Slater determinant wave function 
\begin{align}
  \psi(x_1,\cdots, x_N)=\frac{1}{\sqrt{N!}}\sum_{\sigma\in S_N}(-)^\sigma\chi_1(x_{\sigma(1)})\cdots \chi_N(x_{\sigma(N)})
\end{align}
can be written as 
\begin{align}
    \ket{\psi}=c_1^\dagger \cdots c_N^\dagger \ket{0}.
\end{align}

Using this representation, we can easily confirm Wick's theorem. 
The two-point function can be computed as 
\begin{align}
\label{2pt:c}
 G(x;y):=   \bra{\psi}c^\dagger(x) c(y)\ket{\psi}=\sum_{i,j\,:\,\text{all}}\chi^\ast_i(x) \chi_j(y)\bra{\psi}c^\dagger_i c_j\ket{\psi}=\sum_{i=1}^N\chi^\ast_i(x) \chi_i(y).
\end{align}
We also have
\begin{align}
    \bra{\psi}c^\dagger_{i_k}\cdots c^\dagger_{i_1} c_{j_1} \cdots c_{j_\ell}\ket{\psi}
    =\delta_{k,\ell}\sum_{\sigma\in S_k}(-)^\sigma \delta_{i_{\sigma(1)}, j_1}\cdots\delta_{i_{\sigma(k)}, j_k}
\end{align}
where $i_1, \cdots, i_k, j_1, \cdots, j_\ell$ are restricted to $1,\cdots,N$.
We thus obtain Wick's contraction rule:
\begin{align}
\label{Wick:G}
    \bra{\psi}c^\dagger(x_k)\cdots c^\dagger(x_1) c(y_1) \cdots c(y_\ell)\ket{\psi}
    =\delta_{k,\ell}\sum_{\sigma\in S_k}(-)^\sigma G(x_{\sigma(1)};y_1)\cdots G(x_{\sigma(k)};y_k).
\end{align}
For example, a four-point function 
\begin{align}
    G_4(x_1,x_2;x_3,x_4):= \bra{\psi} c^\dagger (x_1)c^\dagger (x_2) c(x_3)c(x_4)\ket{\psi}
\end{align}
is computed as 
\begin{align}
   G_4(x_1,x_2;x_3,x_4)&=\sum_i \chi_{i}^\ast(x_1) \chi_{i}(x_4) \sum_j \chi_{j}^\ast(x_2)\chi_{j}(x_3)
     -\sum_i \chi_{i}^\ast(x_1) \chi_{i}(x_3) \sum_j \chi_{j}^\ast(x_2)\chi_{j}(x_4)
     \nn
     &=G(x_1;x_4)G(x_2;x_3)-G(x_1;x_3)G(x_2;x_4).
\end{align}

We now show that the reduced density matrix on $A$ effectively takes a form like \eqref{dens:indep}. 
To show this, it is useful to introduce effective ladder operators in the subregion $A$ as 
\begin{align}
    c_\text{eff}(x)=\sum_{i\,:\,\text{all}}\psi_i(x) d_i,\qquad  (x\in A).
\end{align}
This $c_\text{eff}(x)$ is different from $c(x)$ in \eqref{mode:c}, but reproduces the same correlation functions as \eqref{2pt:c}, \eqref{Wick:G} for the state $\ket{\psi}$. 
Here $d^\dagger_i, d_i$ are fermionic ladder operators satisfying $\{d_i ,d^\dagger_j\}=\delta_{i,j}$, and 
$\psi_i(x)$ are orthonormal wave functions in region $A$ as
\begin{align}
    \int_A dx\, \psi_i(x) \psi^\ast_j(x)=\delta_{ij}, \qquad  \sum_{i\,:\,\text{all}}\psi_i(x) \psi^\ast_i(y)=\delta(x,y).
\end{align}
In particular, $\psi_i(x)$ with $i=1,\cdots,N$ are given by
\begin{align}
    \psi_i(x)=\frac{1}{\sqrt{\lambda_i}}\tilde{\chi}_i(x),
\end{align}
\text{i.e.}, $\psi_i(x)\, (i=1,\cdots,N)$ are the wave functions defined in \eqref{k-wf} with $I=\{i\}$.

Using the ladder operators $d^\dagger_i, d_i$,  we can represent the reduced density matrix on  $A$ as 
\begin{align}
\label{dm:2nd}
    \rho_A^\mathrm{2nd}=\prod_{i=1}^N \frac{e^{-\epsilon_i d^\dagger_i d_i}}{1+e^{-\epsilon_i}}
\end{align}
with
\begin{align}
    \epsilon_i= \log \frac{1-\lambda_i}{\lambda_i}.
\end{align}
This is the realization of \eqref{dens:indep}. 
It is easy to confirm that this reduced density matrix reproduces the (R\'enyi) entanglement entropy \eqref{formula_EE} and \eqref{formula_Renyi}.
The form \eqref{dm:2nd} means that the modular Hamiltonian is quadratic
\begin{align}
    K=\sum_{i=1}^N \epsilon_i d^\dagger_i d_i + \text{(const.)},
\end{align}
and the entanglement spectrum is $\{\epsilon_1,\cdots,\epsilon_N\}$.
This also explains that the effective dimension of the reduced Hilbert space is $2^N$ and thus the maximum entropy is $N \log 2$. 
Note also that we have
\begin{align}
    \lambda_i = \frac{e^{-\epsilon_i}}{1+e^{-\epsilon_i}}.
\end{align}

Let us now confirm that the reduced density matrix \eqref{dm:2nd} reproduces the correct correlation functions \eqref{2pt:c}, \eqref{Wick:G} on subregion $A$.
First, the two-point function is computed as 
\begin{align}
    \tr [c_\text{eff}^\dagger(x)c_\text{eff}(y) \rho_A^\mathrm{2nd}]=\sum_{i}\lambda_i \psi^\ast_i(x)\psi_i(y)=\sum_i \tilde{\chi}_i^\ast(x) \tilde{\chi}_i(y)=\sum_{i=1}^N\chi^\ast_i(x) \chi_i(y)=G(x;y)
\end{align}
for $x,y\in A$.
Next, we have 
\begin{align}
       \tr [ d^\dagger_{i_k}\cdots d^\dagger_{i_1} d_{j_1} \cdots d_{j_\ell}\rho_A^\mathrm{2nd}]
       =
       \delta_{k,\ell}\sum_{\sigma\in S_k}(-)^\sigma \delta_{i_{\sigma(1)}, j_1}\cdots\delta_{i_{\sigma(k)}, j_k}
       \lambda_{i_1}\cdots \lambda_{i_k},
\end{align}
where $i_1, \cdots, i_k, j_1, \cdots, j_\ell$ are restricted to $1,\cdots,N$.
It leads to 
\begin{align}
   \tr [ c_\text{eff}^\dagger(x_k)\cdots c_\text{eff}^\dagger(x_1) c_\text{eff}(y_1) \cdots c_\text{eff}(y_\ell)\rho_A^\mathrm{2nd}]
       &=  \delta_{k,\ell}\sum_{\sigma\in S_k}(-)^\sigma \left[G(x_{\sigma(1)};y_1)\cdots G(x_{\sigma(k)};y_k)\right]
       \nn
       &=\bra{\psi}c^\dagger(x_k)\cdots c^\dagger(x_1) c(y_1) \cdots c(y_\ell)\ket{\psi}.
\end{align}

$p_k$ and $\rho_{k,A}$ appeared in the first-quantized picture can also be obtained as follows. 
In the Fock space with respect to the ladder operators $d_i, d^\dagger_i$, 
$k$-particle states are spanned by 
\begin{align}
    \ket{I}:=\prod_{i\in I} d^\dagger_i \ket{0}, \qquad I \in F_k.
\end{align}
For example, for $I=\{1,2\}$, we have $\ket{I}=d^\dagger_1 d^\dagger_2\ket{0}$.
In this basis, the reduced density matrix is diagonal with the following components:
\begin{align}
    \bra{I}\rho_A^\mathrm{2nd} \ket{J}
    =\left(\prod_{i\in I}\frac{e^{-\epsilon_i}}{1+e^{-\epsilon_i}}\right)\left( \prod_{j\in \bar{I}}\frac{1}{1+e^{-\epsilon_j}}\right)\delta_{I,J}
    =\lambda_I \bar{\lambda}_I \delta_{I,J}.
\end{align}

We define the identity operator $\hat{1}_k$ on the $k$-particle space as
\begin{align}
    \hat{1}_k:=\sum_{I\in  F_k}\ketbra{I}{I}.
\end{align}
The probabilities $p_k$ that we find $k$ particles in region $A$ are given by
\begin{align}
    p_k=\tr [\hat{1}_k\,\rho_A^\mathrm{2nd}]=\sum_{I\in  F_k}\lambda_I \bar{\lambda}_I.
\end{align}
$\rho_A^\mathrm{2nd}$ can be decomposed into $k$-particle sectors as 
\begin{align}
    \rho_A^\mathrm{2nd} =\sum_{k=0}^N p_k \rho_{k,A}^\mathrm{2nd}, 
\end{align}
where
\begin{align}
    \rho_{k,A}^\mathrm{2nd}:=\frac{1}{p_k} \sum_{I \in F_k}\lambda_I \bar{\lambda}_I\ketbra{I}{I}.
\end{align}

We now compute the matrix elements of $\rho_{k,A}^\mathrm{2nd}$ and they agree with \eqref{elem:kbody-A}.
We define position-basis states in the second-quantized picture as
\begin{align}
    \ket{y_1,\cdots, y_k}:=\frac{1}{\sqrt{k!}}c_\text{eff}^\dagger(y_1)\cdots c_\text{eff}^\dagger(y_k)\ket{0}
    =\frac{1}{\sqrt{k!}}\sum_{i_1, \cdots, i_k:\text{all}}\psi^\ast_{i_1}(y_1)\cdots \psi^\ast_{i_k}(y_k)d^\dagger_{i_1}\cdots d^\dagger_{i_k}\ket{0}.
\end{align}
When we compute matrix elements of $\rho_{k,A}^\mathrm{2nd}$, we can restrict the sum of indices $i_j$ onto $1,\cdots, N$ as
\begin{align}
    \ket{y_1,\cdots, y_k}\to &\frac{1}{\sqrt{k!}}\sum_{i_1, \cdots, i_k=1}^N\psi^\ast_{i_1}(y_1)\cdots \psi^\ast_{i_k}(y_k)d^\dagger_{i_1}\cdots d^\dagger_{i_k}\ket{0}
    =\sum_{I \in F_k}\psi^\ast_I(\vy) \ket{I},
\end{align}
where $\psi_I(\vy)$ are the $k$-body wave functions defined in \eqref{k-wf}.
Thus, the matrix elements of $\rho_{k,A}^\mathrm{2nd}$ in the position-basis are obtained as
\begin{align}
    \mel{\vy}{\rho_{k,A}^\mathrm{2nd}}{\vyp}=\frac{1}{p_k} \sum_{I \in F_k}\lambda_I \bar{\lambda}_I \psi_I(\vy)\psi^\ast_I(\vyp),
\end{align}
which agree with \eqref{elem:kbody-A}.


\section{Interval with Dirichlet conditions}
\label{app:diric}
In section~\ref{sec:free-1dim},  the periodic boundary condition is imposed on particles. 
Instead, we consider a finite interval $-L/2\leq x \leq L/2$
and impose the Dirichlet boundary condition $\psi(\pm L/2)=0$.
The eigenfunctions of a single particle  are
\begin{align}
    \chi_n(x)=\frac{1}{\sqrt{2L}i^{n-1}} \left[e^{i\frac{n \pi x}{L}}+(-1)^{n-1}e^{-i\frac{n \pi x}{L}}\right],  \qquad (n=1,2,3, \cdots).
\end{align}

\begin{figure}[H]
\centering
\includegraphics[width=10cm]{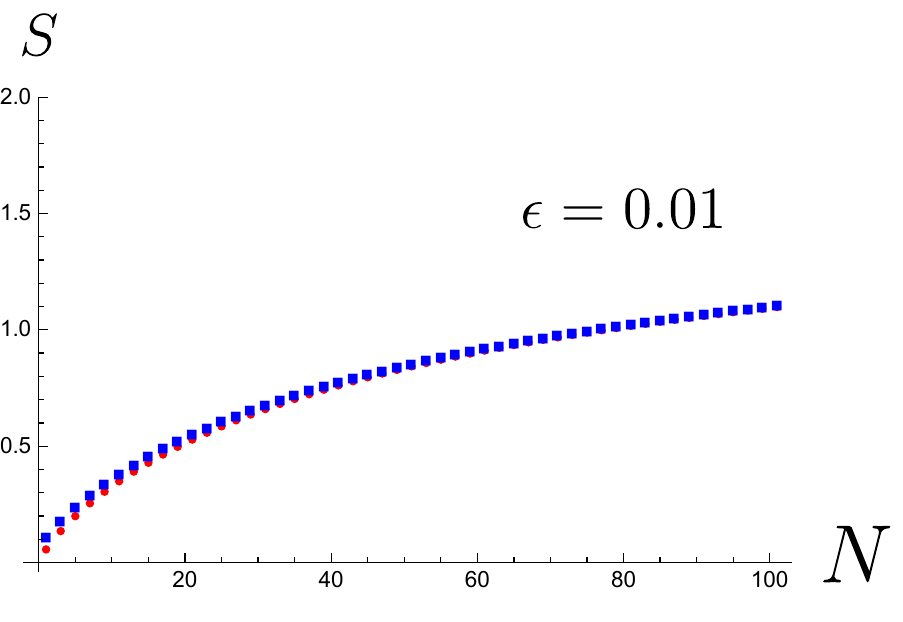}
\caption{Entanglement entropy for an interval. 
The subregion is an interval $A=(-\epsilon L/2, \epsilon L/2)$ with $\epsilon=0.01$. 
The blue square markers are the entanglement entropy for the Dirichlet boundary condition. The red round markers are for the periodic condition.}
\label{fig:box}
\end{figure}

The (R\'enyi) entanglement entropy for a subregion $A$ in the interval can be computed by evaluating the eigenvalues of the overlap matrix $X(A)$ in \eqref{def:X} for these wave functions. 
We show the entanglement entropy 
for a single interval $A=(-\epsilon L/2, \epsilon L/2)$ with $\epsilon=0.01$ in Fig.\ref{fig:box}. 
The result is similar to that for the periodic boundary condition. 
The results imply that effects of the boundary conditions are negligible for small intervals, as expected. 

\begin{figure}[H]
\centering
\includegraphics[width=10cm]{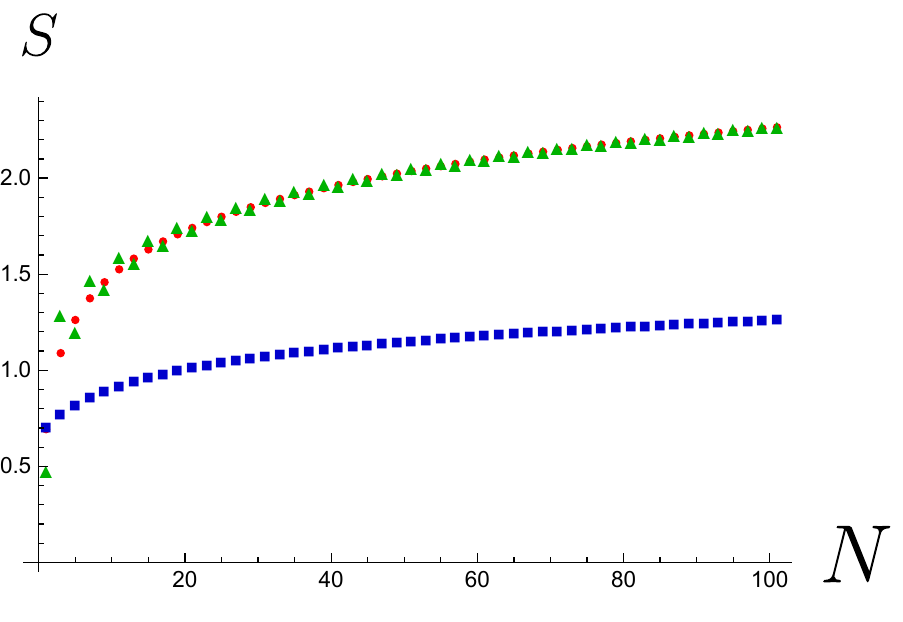}
\caption{Entanglement entropy for a half region. 
The blue square and green triangular markers are the entanglement entropies 
for the left half subregion $A=(-L/2,0)$ and the middle half subregion $A=(-L/4,L/4)$, respectively, 
in the Dirichlet boundary condition. The red round markers are for the half region in the circle (the periodic condition).
The green triangular markers are almost the same as the red round ones for large $N$. 
}
\label{fig:halfcomp}
\end{figure}

The independence of boundary conditions holds even for the half region as shown in Fig.~\ref{fig:halfcomp}.
The entanglement entropy for the middle half region, $A=(-L/4,L/4)$, with the Dirichlet boundary condition at $x=\pm L/2$ is almost the same at large $N$ as that for the half region on the circle. 
However, if the half region is attached to the boundary, \textit{e.g.} $A=(-L/2,0)$, the behavior of the entanglement entropy is different. 
In this case, the degrees of freedom at the edge $x=- L/2$ of the subregion are frozen because of the Dirichlet boundary condition. 
The entanglement entropy behaves as if the subregion  has only one entangling surface (edge). 
Thus, the entropy is smaller than that for the middle subregion which has two entangling surfaces. 

\section{Large $N$ computations for entropy of two intervals via the Fisher-Hartwig conjecture}
\label{app:FH}
In this appendix, we derive the large $N$ result \eqref{twointEE} of the (R\'enyi) entanglement entropy for two intervals.

We have shown that if we know  the eigenvalues $\lambda_i$ of the overlap matrix $X$, we can compute the R\'enyi entropy as
\begin{align}
    S^{(n)}=\sum_{i=1}^N H^{(n)}(\lambda_i)
\end{align}
with  $H^{(n)}(\lambda)=\frac{\log[\lambda^n+(1-\lambda)^n]}{1-n}$.
However, without directly diagonalizing the matrix $X$, we can compute the entropy by a method developed in \cite{Jin_2004}.
We first introduce the following matrix with a parameter $z$:
\begin{align}
\label{def:Tmat}
    T(z):=z\,I-X.
\end{align}
Then, the R\'enyi entropy can  be expressed in terms of the determinant $\det T(z)=\prod_{i=1}^N(z-\lambda_i)$ as
\begin{align}
\label{renyi-T}
    S^{(n)}=\frac{1}{2\pi i}\oint d z\, H^{(n)}\left(z\right)\frac{d \log \det T(z)}{dz},
\end{align}
where the integral contour encircles the interval $[0,1]$. 
Thus, our task is to compute the determinant $\det T(z)$.

If the overlap matrix $X$ is a Toeplitz matrix such that the components $X_{jk}$ depend only on the difference $(j-k)$, the matrix $T(z)$ is too. 
The Fisher-Hartwig conjecture \cite{wu1966theory, fisher1969toeplitz} predicts the determinant of a Toeplitz matrix at large $N$.
For free $N$ fermions on the circle where each energy eigenfunctions are given by \eqref{free_chi}, the overlap matrix $X$ for any numbers of disjoint intervals can be written as a Toeplitz matrix as follows.
Supposing that $N$ is an odd number as $N=2K+1$,  the set of $N$ energy eigenfunctions are 
\begin{align}
   \left\{ \varphi_j(x)=\frac{1}{\sqrt{L}}e^{\frac{2\pi i}{L}j x}\, \Big{|}\, j=-K, \cdots, K\right\}.
\end{align}
In this basis, the overlap matrix for a single interval
$I_1=((-d-r) L/2, (-d+r) L/2)$ is
\begin{align}
\label{appc:Xsing}
    X_{jk}(I_1)=\int^{(-d+r) L/2}_{(-d-r) L/2}\!\!\!dx\varphi_j(x)\varphi_k^\ast(x)=\frac{e^{-\pi i(j-k)d}}{\pi(j-k)}\sin [\pi(j-k)r],
\end{align}
which is a Toeplitz matrix because the $(j,k)$ elements depend only on $j-k$.
Since the sum of Toeplitz matrices is also a Toeplitz matrix, the overlap matrix for the union of disjoint intervals is a Toeplitz matrix.
In particular, 
for two intervals
$I_1=((-d-r) L/2, (-d+r) L/2)$ and $I_2=((d-r) L/2, (d+r) L/2)$, 
the overlap matrix is given by the following Toeplitz matrix 
\begin{align}
\label{X:twoint}
    X_{jk}(I_1 \cup I_2)=\frac{1}{\pi(j-k)}\left\{\sin [\pi(j-k)(d+r)]-\sin [\pi(j-k)(d-r)]\right\}.
\end{align}

We now see the large $N$ behavior of $\det T(z)$ for $T(z)=z- X(I_1 \cup I_2)$. Since the component $T_{jk}$ depends only on $j-k$, we write it as $t_{j-k}:=T_{jk}$. 
Let's consider the Fourier transform of $t_{n},\, (n=0, \cdots, N-1)$;
\begin{align}
    t_{n}=\int^{\pi}_{-\pi}\frac{d\theta}{2\pi} e^{in\theta}t(\theta).
\end{align}
The function $t(\theta)$ has four discontinuities located at $\pm \theta_\pm$ with $\theta_\pm=\pi(d\pm r)$ in the range $[-\pi,\pi]$ as 
\begin{align}
    t(\theta)=
    \begin{cases}
    z, & \theta \in [-\pi, -\theta_+] \cup [\theta_+,\pi]\\
    z-1, & \theta \in [-\theta_+, -\theta_-] \cup [\theta_-,\theta_+]\\
    z, & \theta \in [-\theta_-, \theta_-]
    \end{cases}.
\end{align}
We also represent the positions of the discontinuities as $\theta_1=-\theta_+, \theta_2=-\theta_-, \theta_3=\theta_-, \theta_4=\theta_+$.
Then, $t(\theta)$ can also be written as 
\begin{align}
    t(\theta)=f_0 \prod_{p=1}^4 e^{i b_p[\theta-\theta_p-\pi \sgn(\theta-\theta_p)]}
\end{align}
where $b_1=-b_4=\beta(z)+m_1, b_2=-b_3=-\beta(z)-m_2$ and 
$f_0=z e^{2i \sum_{j=1,2}\theta_j b_j}$ with 
\begin{align}
    \beta(z)=\frac{1}{2\pi i}\log \frac{z}{z-1}
\end{align}
and $m_1, m_2$ are arbitrary integers.

The Fisher-Hartwig conjecture states that the large $N$ asymptotic behavior of $\det T$ is given by\footnote{Here we sum over the inequivalent  solutions, namely, sum over $m_1,m_2$, following  \cite{Calabrese_2010}.}
\begin{align}
\label{FH_detT}
    \det T \simeq \sum_{m_1,m_2 \in \mathbb{Z}}f_0^N N^{-2b_1^2-2b_2^2}\prod_{i=1,2}[G(1+b_i)G(1-b_i)]^2
    \prod_{1\leq p \neq q \leq 4}(1-e^{i(\theta_p-\theta_q)})^{b_p b_q},
\end{align}
where $G$ is the Barnes $G$-function defined as
\begin{align}
    G(1+b)=(2 \pi)^{b / 2} \exp \left(-\frac{b+b^{2}(1+\gamma)}{2}\right) \prod_{k=1}^{\infty}\left\{\left(1+\frac{b}{k}\right)^{k} \exp \left(\frac{b^{2}}{2 k}-b\right)\right\}.
\end{align}

We can show 
\begin{align}
    \mathrm{Re}[(\beta(z)+m)^2]>\mathrm{Re}[\beta(z)^2]
\end{align}
for arbitrary nonzero integers $m$ 
on the contours $z=x\pm i \epsilon$ ($-1<x<1$) with small but nonzero $\epsilon>0$.
Thus, $m_1=m_2=0$ is the leading contribution in the sum in  \eqref{FH_detT}, and we obtain 
\begin{align}
     &\det T 
    \simeq
     e^{2i N (\theta_1-\theta_2)\beta} 
     [4N^2 \sin(\pi (d+r)) \sin(\pi (d-r))]^{-2 \beta^2} 
     \left[\frac{\sin(\pi r)}{\sin(\pi d)}\right]^{-4\beta^2}
     [G(1+\beta)G(1-\beta)]^4.
\end{align}

Inserting this equation into \eqref{renyi-T}, the R\'enyi entropy is
\begin{align}
        S^{(n)}(I_1\cup I_2)\simeq N a_0^{(n)} + a_1^{(n)} A_N(d,r)+2\Upsilon_n,
\end{align}
where
\begin{align}
    &a_0^{(n)}=\frac{N(\theta_1-\theta_2)}{\pi}\oint dz H^{(n)}(z)\frac{d\beta}{dz}= 0,
    \\
    &a_1^{(n)}=-\frac{1}{\pi i}\oint dz H^{(n)}(z)\frac{d\beta^2}{dz},
    \\
   & A_N(d,r)=\log  [2N \sin(\pi (d+r))]+\log  [2N \sin(\pi (d-r))]+2 \log\left[\frac{\sin(\pi r)}{\sin(\pi d)}\right],
   \\
   &\Upsilon_n=\frac{1}{\pi i}\oint dz H^{(n)}(z)\frac{d \log [G(1+\beta)G(1-\beta)] }{dz}.
\end{align}
The integrals also appear in the single interval case (see \cite{Jin_2004,  Bonsignori:2019naz}), and can be simplified as
\begin{align}
    &a_1^{(n)}=
    \frac{1}{\pi^2}\oint dz H^{(n)}(z)\frac{\beta(z)}{z(1-z)}
    =\int^{\infty}_{-\infty}dw  \frac{2nw}{(n-1)}[\tanh(\pi n w)-\tanh(\pi w)]=\frac{1}{6}\left(1+\frac{1}{n}\right),
    \\
   &\Upsilon_n=\frac{n
   }{i(1-n)}\int^{\infty}_{-\infty}dw [\tanh(\pi n w)-\tanh(\pi w)]\log \frac{\Gamma\left(\frac{1}{2}+i w\right)}{\Gamma\left(\frac{1}{2}-i w\right)}.
\end{align}
Therefore, we obtain 
\begin{align}
        S^{(n)}(I_1\cup I_2)\simeq  \frac{1}{6}\left(1+\frac{1}{n}\right)\left[2\log[2N\sin (\pi r)]+\log\frac{ \sin[\pi (d+r)] \sin[\pi (d-r)]}{\sin^2(\pi d)}\right] +2\Upsilon_n.
\end{align}

\newpage
\bibliographystyle{utphys}
\bibliography{ref}
\end{document}